\documentclass[aps,preprint,superscriptaddress,showpacs,preprintnumbers,amsmath,amssymb]{revtex4-1}
\usepackage[colorlinks=true, pdfstartview=FitV, linkcolor=red, citecolor=blue, urlcolor=black, pdftitle={Dispersion relations of Nambu-Goldstone modes at finite temperature and density},pdfauthor={Tomoya Hayata, Yoshimasa Hidaka},pdfsubject={}, pdfkeywords={Nambu-Goldstone theorem, nonrelativistic systems, finite temperature}]{hyperref}

\usepackage{amsmath,amssymb,bm}
\usepackage[pdftex]{graphicx}
\newcommand{\tr}{\mathop{\mathrm{tr}}}

\newcommand{\average}[1]{\langle#1\rangle}
\newcommand{\noise}{\hat{R}}
\newcommand{\memory}{K}
\newcommand{\Lv}{\hat{\mathcal{L}}}
\newcommand{\rank}{\mathop{\mathrm{rank}}}

\newcommand{\NBS}{N_\text{BS}}

\newcommand{\NNG}{N_\text{NG}}

\newcommand{\cP}{\hat{\mathcal{P}}}
\newcommand{\cQ}{\hat{\mathcal{Q}}}

\newcommand{\hphi}{\hat{\phi}}
\newcommand{\hn}{\hat{n}}

\newcommand{\hPhi}{\hat{\Phi}}

\newcommand{\OP}{h}
\newcommand{\hA}{\hat{A}}

\newcommand{\hH}{\hat{H}}

\newcommand{\hO}{\hat{\mathcal{O}}}
\newcommand{\hN}{\hat{N}}

\newcommand{\hQ}{\hat{Q}}
\newcommand{\hT}{\hat{T}}
\newcommand{\hV}{\hat{V}}
\newcommand{\hK}{\hat{K}}
\newcommand{\hpi}{\hat{\pi}}
\newcommand{\hrho}{\hat{\rho}}
\newcommand{\typeA}{{\rm A}}
\newcommand{\typeB}{{\rm B}}
\newcommand{\stiffness}{\rho}
\newcommand{\innerProd}[1]{\bm{(}#1\bm{)}}

\newcommand{\deltaB}{\delta}
\begin{document}

\preprint{RIKEN-QHP-159}
\author{Tomoya Hayata}
\affiliation{Department of Physics, The University of Tokyo, Tokyo 113-0031, Japan}
\affiliation{Theoretical Research Division, Nishina Center, RIKEN, %
             Wako 351-0198, Japan}
\author{Yoshimasa Hidaka}
\affiliation{Theoretical Research Division, Nishina Center, RIKEN, %
             Wako 351-0198, Japan}
\title{Dispersion relations of Nambu-Goldstone modes at finite temperature and density}
\begin{abstract}
We  discuss the dispersion relations of Nambu-Goldstone (NG) modes associated with spontaneous breaking of internal symmetries
at finite temperature and/or density. 
We show that the dispersion relations of type-A (I) and type-B (II) NG modes are linear and quadratic in momentum,
whose imaginary parts are quadratic and quartic, respectively.
In both cases, the real parts of the dispersion relations are larger than the imaginary parts when the momentum is small, so that the 
NG modes can propagate far away. We derive the gap formula for NG modes in the presence of a small explicit breaking term.
We also discuss the gapped partners of type-B NG modes, when the expectation values of a charge density and a local operator that break the same symmetry
 coexist.
\end{abstract}
\maketitle
\section{Introduction} \label{sec:introduction}
The low-energy or long-distance behavior of many body systems is determined by collective excitation modes with zero or almost zero gap. 
The dynamical degrees of freedom, which represent such low-energy excitations are called ``slow variables.''
If a global symmetry is spontaneously broken, it is necessarily accompanied by a slow variable called ``elastic variable," which is defined as the flat direction of the free energy~\cite{Chaikin}:
Because of the infinite degeneracy of thermal states, a continuous transformation under the broken symmetry, labeled by the elastic variable $\pi$, 
does not cost the free energy. 
The free energy increases with $\big(\partial_i\pi(\bm{x})\big)^2$ by a slow variation of $\pi(\bm{x})$ in space. 

When a symmetry is spontaneously broken, the expectation value of commutation relation between the broken charge operator $\hQ_a$ and the elastic variable $\hpi_i(\bm{x})$ does not vanish, i.e.,
\begin{equation}
\average{[i\hQ_a,\hpi_i(\bm{x})]}\neq0, \label{eq:SSB}
\end{equation}
where $\average{\cdots}$ denotes the expectation value. 
This implies that  the elastic variable couples to the broken charge in the way of canonical pairs, and then they form a gapless propagating mode, i.e., the Nambu-Goldstone (NG) mode (if the mode is quantized, it is called the NG boson)~\cite{Nambu:1961tp,Goldstone:1961eq,Goldstone:1962es}.

To understand the low-energy physics associated with spontaneous symmetry breaking, the general relation between the number of broken symmetries, elastic variables, and NG modes has been actively investigated~\cite{Brauner:2010wm}.
In the case of spontaneous breaking of internal symmetries,  
the number of independent-elastic variables is equal to the number of broken symmetries (or equivalently generators), $\NBS$.
However, this is not always true for spontaneous breaking of spacetime symmetries. The number of elastic variables is equal or smaller than $\NBS$~\cite{Hayata:2013vfa} (see discussions for spontaneous breaking of spacetime symmetries~\cite{Volkov:1973vd,Ogievetsky,Ivanov:1975zq,Low:2001bw, Watanabe:2013iia,Nicolis:2013sga,Endlich:2013vfa,Brauner:2014aha,Watanabe:2014zza,Goon:2014ika}).

For the Lorentz invariant system, the number of independent NG modes coincides with $\NBS$~\cite{Goldstone:1962es}.
On the other hand, when the system is not Lorentz invariant, the number of them is not necessarily equal to $\NBS$.
For internal symmetry breaking,  Nielsen and Chadha~\cite{Nielsen:1975hm} classified the NG modes using their dispersion relations into two types: type-I (II) NG mode whose energy is proportional to odd (even) powers of momentum. They showed the inequality $N_{\text{I}} + 2N_{\text{II}}\geq N_{\text{BS}}$, where $N_{\text{I}}$ and $N_{\text{II}}$ are the number of type-I and type-II NG modes, respectively. 

Schafer et al. pointed out the importance of the commutation relation between broken charges, and showed that if all $\average{[i\hQ_a,\hQ_b]}$ vanish, the number of NG modes coincides with $\NBS$~\cite{Schafer:2001bq}. 
The relation between the nonvanishing expectation value of the charge density and the existence of type-II NG modes was discussed using the effective Lagrangian approach by Leutwyler~\cite{Leutwyler:1993gf}. 
Later, Nambu discussed that if $\average{[i\hQ_a,\hQ_b]}\neq0$, $\hQ_a$ and $\hQ_b$ are not independent in the sense of canonical variables, and thus it reduces the independent propagation of NG modes~\cite{Nambu:2004}. 
 
Recently, the classification using $\average{[i\hQ_a,\hQ_b]}$ has been discussed~\cite{Watanabe:2011ec,*Watanabe:2011dk,Watanabe:2012hr,*Watanabe:2014fva,Hidaka:2012ym}.
The NG modes characterized by the nonvanishing $\average{[i\hQ_a,\hQ_b]}$ are classified as type-B NG modes, whose number is given, using the rank of $\average{[i\hQ_a,\hQ_b]}$, by $N_\typeB=\rank\average{[i\hQ_a,\hQ_b]}/2$.
The other NG modes are classified as type-A NG modes, whose number is $N_\typeA=N_\text{BS}-2N_\typeB$~\footnote{In \cite{Hidaka:2012ym}, type-A and type-B NG modes are called type-I and type-II NG modes because they usually coincide with the Nielsen-Chadha classification.}. 
Since the total number of NG modes is $\NNG=N_\typeA+N_\typeB$,  the following counting rule holds:
\begin{equation}
\NNG= \NBS- \frac{1}{2}\rank \average{[i\hQ_a,\hQ_b]}.
\label{eq:NGRelation}
\end{equation}
This equality was conjectured by Watanabe and Brauner~\cite{Watanabe:2011ec,*Watanabe:2011dk},
and later proved by Watanabe and Murayama, and independently one of the authors~\cite{Watanabe:2012hr,Hidaka:2012ym}.
For typical cases, type-A and type-B NG modes coincide with type-I and type-II NG modes, respectively.
Strictly speaking, the broken charges are not well defined at the infinite volume limit, so that $\average{[i\hQ_a,\hQ_b]}$ is not.
Therefore, $\average{[i\hQ_a,\hQ_b]}$ in Eq.~\eqref{eq:NGRelation} should be understood as
\begin{equation}
\begin{split}
\lim_{V\to\infty}\frac{1}{V}\int_V d^3x \average{[i\hQ_a,\hn_b(\bm{x})]},
\end{split}
\end{equation}
where $V$ is the volume of the system and $\average{[i\hQ_a,\hn_b(\bm{x})]}$ is well defined.
Equation \eqref{eq:NGRelation} was first discussed for spontaneous breaking of internal symmetries; however, it seems to
be satisfied for several systems in which spacetime symmetry is spontaneous broken~\cite{Watanabe:2014pea,Kobayashi:2014xua,*Kobayashi:2014eqa}.
Another counting rule to cover both spontaneous breaking of internal and spacetime symmetries was proposed on the basis of the Bogoliubov theory~\cite{Takahashi:2014vua}.

In this paper, we focus on spontaneous breaking of internal symmetries, and 
generalize the argument for the dispersion relations of type-A and type-B NG modes~\cite{Hidaka:2012ym}
into systems at finite temperature and/or density. 
Previous works were mostly limited at zero or low temperatures. 
For this purpose, we employ generalized Langevin equations for the slow variables, which are formally obtained using the projection operator method~\cite{Mori}.
We show that the dispersion relations of type-A and B NG modes have the forms of $\omega = v_\typeA k-i\Gamma_\typeA k^2$ and  $\omega = v_\typeB k^2-i\Gamma_\typeB k^4$,
respectively, where $v_{\typeA,\typeB}$ and $\Gamma_{\typeA,\typeB}$ are some constants that 
depend on the detail of systems, not only on the symmetry breaking pattern.
Our method is applicable to not only low temperature where  an effective Lagrangian method works~\cite{Coleman:1969sm,*Callan:1969sn,Gasser:1983yg,*Gasser:1984gg,Leutwyler:1993gf,Watanabe:2012hr,*Watanabe:2014fva,Brauner:2014ata,Andersen:2014ywa} but also higher temperature where heavy degrees of freedom are excited, if the system is still in a broken phase. We also discuss the existence of gapped partners for type-B NG modes~\cite{Kapustin:2012cr,Gongyo:2014sra}. We show that the gapped partners appear
when nonvanishing expectation values $\average{[i\hQ_a,\hphi_i(\bm{x})]}\neq0$ and $\average{[i\hQ_a,\hn_b(\bm{x})]}\neq0$ coexist, where $\hphi_i(\bm{x})$ is an local operator  that is not a charge density. As is discussed later, this coexistence leads to the mixing of $\hphi_i(\bm{x})$ and $\hn_a(\bm{x})$ in the equations of motion and thus they create gapless and gapped modes.

This paper is organized as follows.
In Sec.~\ref{sec:Examples}, we discuss the dispersion relations for type-A and type-B NG modes at finite temperature using simple classical models.
In Sec.~\ref{sec:SSB}, the relation between elastic variables and broken symmetries is summarized.
In Sec.~\ref{sec:LangevinEquation}, we review Mori's projection operator method that is used for deriving the dispersion relations of NG modes.
In Sec.~\ref{sec:DispersionRelation}, we discuss the dispersion relations of NG modes at finite temperature, the existence of gapped partners, and the gap formula when a small explicit breaking term is added into the Hamiltonian. 
We also discuss the mixing between the type-A or type-B NG and hydrodynamic modes.  
Section~\ref{sec:summary} is devoted to a summary.

\section{Dispersion relations of Nambu-Goldstone modes in simple Langevin systems}\label{sec:Examples}
Before a detailed analysis, we study the dispersion relations of type-A and type-B NG modes at finite temperature using simple classical models.
At finite temperature, the NG mode can couple to the hydrodynamic mode. Such a mixing does modify the dispersion relation of the NG mode, but does not modify the powers of it.
The effect of the mixing will be discussed in Sec.~\ref{sec:HydrodynamicModes}. For simplicity, here, we do not treat it. 

First, we consider an example of the type-A NG mode. We suppose that the $U(1)$ symmetry is spontaneously broken.
We write the $U(1)$ charge and its density as $Q$ and $n(\bm{x})$, respectively. 
The symmetry breaking implies that there exists an elastic variable $\pi(\bm{x})$ such that 
\begin{equation}
\begin{split}
\{\pi(\bm{x}), Q\}_P \equiv 1\neq0 , \label{eq:ConjugateA}
\end{split}
\end{equation}
where $\{ \ ,\  \}_P$ denotes the Poisson bracket.  The elastic variable $\pi(\bm{x})$ is not a charge density, so that the NG mode belongs to the type-A NG modes.
The Poisson bracket between $\pi(\bm{x})$ and $n(\bm{x}')$, $\{\pi(\bm{x}),n(\bm{x}')\}_P$, is local, i.e., it is proportional to the delta function. In order to satisfy Eq.~\eqref{eq:ConjugateA}, it  reads
\begin{equation}
\begin{split}
\{\pi(\bm{x}),n(\bm{x}')\}_P=\delta^{(3)}(\bm{x}-\bm{x}'). \label{eq:PoissonBracket1}
\end{split}
\end{equation}
Strictly speaking, total derivative terms such as $\partial_i^2\delta^{(3)}(\bm{x}-\bm{x}')$ that vanish in Eq.~\eqref{eq:ConjugateA} may appear in the right-hand side of Eq.~\eqref{eq:PoissonBracket1}. However, such terms do not contribute to the dispersion relations of NG modes in the leading order of the derivative expansion.
Therefore, we do not take into account the derivative terms.

The free energy can be given as
\begin{equation}
\begin{split}
F[n,\pi] =\int d^3x\Bigl( \frac{1}{2}\chi^{-1}n(\bm{x})n(\bm{x}) + \frac{\rho}{2}\partial_i\pi(\bm{x})\partial_i\pi(\bm{x})\Bigr) + \cdots,
\label{eq:ElasticFreenergy}
\end{split}
\end{equation}
where $\cdots$ denotes higher-order derivative and nonlinear terms.
There is no $\pi^2$ term that does not contain derivatives because a constant change of $\pi$ does not cost the free energy. 
In contrast, the susceptibility $\chi$, which is defined by
\begin{equation}
\begin{split}
\chi \equiv \frac{1}{V}\int d^3x d^3x'\average{n(\bm{x})n(\bm{x}')},
\end{split}
\end{equation}
 is generally finite. Here, we assumed $\average{n(\bm{x})}=0$. For the case with $\average{n(\bm{x})}\neq0$, one may use $\delta n(\bm{x})=n(\bm{x})-\average{n(\bm{x})}$ as the degrees of freedom instead of $n(\bm{x})$.

We aim to derive the dispersion relations of NG modes at finite temperature. 
For this purpose, it is useful to introduce the Langevin equations, which describe slow motions of $\pi(t,\bm{x})$ and $n(t,\bm{x})$.
Their equations are written as
\begin{align}
\partial_0 \pi(t,\bm{x}) &= \{\pi(t,\bm{x}), F\}_P - \gamma\frac{\partial F}{\partial \pi(t,\bm{x})} + \xi_\pi(t,\bm{x}),\\
\partial_0 n(t,\bm{x}) &= \{n(t,\bm{x}), F\}_P + \sigma\partial_i^2\frac{\partial F}{\partial n(t,\bm{x})} + \xi_n(t,\bm{x}). \label{eq:EOMn}
\end{align}
The first, second, and third terms in the right-hand sides denote 
the streaming, dissipation (friction), and noise terms, respectively. The noises satisfy the so-called fluctuation-dissipation theorem:
\begin{align}
\average{\xi_\pi(t,\bm{x})\xi_\pi(t',\bm{x}')} &= 2T\gamma \delta(t-t')\delta^{(3)}(\bm{x}-\bm{x}'),\\
\average{\xi_n(t,\bm{x})\xi_n(t',\bm{x}')} &= -2T\sigma \partial_i^2\delta(t-t') \delta^{(3)}(\bm{x}-\bm{x}'), \label{eq:FDT2}
\end{align}
where $\gamma$ and $\sigma$ are the diffusion parameter and the transport coefficient, respectively. 
The spatial derivative term in Eq.~\eqref{eq:EOMn} results from the conservation law:
From Fick's law, the dissipative part of the current behaves like
\begin{equation}
\begin{split}
j^i(t,\bm{x}) = D \partial^i n(t,\bm{x}). \label{eq:FicksLaw}
\end{split}
\end{equation}
Therefore, the continuity equation leads to $\partial_0 n(t, \bm{x})=-\partial_i j^i(t,\bm{x})=D\partial_i^2 n(t,\bm{x})$, which reproduces
the derivative term in Eq.~\eqref{eq:EOMn} with $D=\sigma\chi^{-1}$.

From the free energy~\eqref{eq:ElasticFreenergy}, we have
\begin{align}
\partial_0\pi(t,\bm{x}) &=  \chi^{-1} n(t,\bm{x}) + \gamma \rho\partial_i^2 \pi(t,\bm{x}) +\xi_\pi(t,\bm{x}),\\
\partial_0n(t,\bm{x}) &= \rho \partial_i^2 \pi(t,\bm{x}) + \sigma \chi^{-1} \partial_i^2 n(t,\bm{x}) +\xi_n(t,\bm{x}).
\end{align}
Since the noise terms are independent of $n(t,\bm{x})$ and $\pi(t,\bm{x})$, they do not contribute to the dispersion relations.
Dropping them, we obtain the equation of motion for $\pi(t,\bm{x})$,
\begin{equation}
\begin{split}
(\partial_0-\sigma \chi^{-1} \partial_i^2 )(\partial_0-\gamma \rho\partial_i^2)\pi(t,\bm{x}) =  \chi^{-1}\rho \partial_i^2 \pi(t,\bm{x}).
\end{split}
\end{equation}
Then, we find the dispersion relation in the leading order of momentum $k=|\bm{k}|$,
\begin{equation}
\begin{split}
\omega = \pm v k -i\Gamma k^2 ,
\end{split}
\label{eq:typeAdispersion}
\end{equation}
where $v=\sqrt{\chi^{-1}\rho}$, and $\Gamma=(\sigma\chi^{-1}+\gamma\rho)/2$, respectively.
Therefore, as long as $v\neq0$, the type-A NG mode is classified as the type-I NG mode, whose imaginary part of the dispersion relation is of order $k^2$.

Next, let us consider an example of the type-B NG mode. Suppose that a global $SU(2)$ symmetry is spontaneously broken into a $U(1)$ symmetry.
Its charge densities are denoted by $n_i(\bm{x})$ ($i=1,2,3$),
which satisfy $\{Q_i, n_j(\bm{x})\}_P= \epsilon_{ijk}n_k(\bm{x})$, where $\epsilon_{ijk}$ is the totally antisymmetric tensor ($\epsilon_{123}=1$).
We consider the situation that $n_3(t,\bm{x})=n_3$ becomes a nonzero constant, i.e., 
\begin{equation}
\begin{split}
\{Q_1,n_2(\bm{x}) \}_P=-\{Q_2,n_1(\bm{x}) \}_P=n_3\neq0. \label{eq:ConjugateB}
\end{split}
\end{equation}
Thus, $Q_1$ and $Q_2$ are the broken charges. 
Equation~\eqref{eq:ConjugateB} implies that their charge densities are canonically conjugate:
\begin{equation}
\begin{split}
\{n_1(\bm{x}),n_2(\bm{x}') \}_P = n_3\delta^{(3)}(\bm{x}-\bm{x}'). \label{eq:PoissonBracketTypeB}
\end{split}
\end{equation}
In this instance, the free energy has the form:
\begin{equation}
\begin{split}
F[n_1,n_2] =\int d^3x\Bigl( \frac{\rho'}{2}\partial_i n_1(\bm{x}) \partial_i n_1(\bm{x}) + \frac{\rho'}{2}\partial_i n_2(\bm{x})\partial_i n_2(\bm{x})\Bigr) + \cdots.
\label{eq:ElasticFreenergyTypeB}
\end{split}
\end{equation}
Unlike the previous example, charge densities, $n_1(\bm{x})$ and $n_2(\bm{x})$, contain the spatial derivative because they are elastic variables.
The number of independent-elastic variables is equal to the number of broken symmetries ($Q_1$ and $Q_2$).
However, they cannot create independent propagating modes because $n_1(\bm{x})$ and $n_2(\bm{x})$ are canonically conjugate.
This can be explicitly seen in the equations of motion:
\begin{align}
\partial_0 n_1(t,\bm{x}) &=  - n_3\rho'\partial_i^2 n_2(t,\bm{x}) - \sigma'\rho'\left(\partial_i^2\right)^2n_1(t,\bm{x}) +\xi_1(t,\bm{x}),\\
\partial_0 n_2(t,\bm{x}) &=  n_3 \rho'\partial_i^2 n_1(t,\bm{x}) - \sigma'\rho'\left(\partial_i^2\right)^2n_2(t,\bm{x})+\xi_2(t,\bm{x}),
\end{align}
where the noises satisfy
\begin{equation}
\begin{split}
\average{\xi_i(t,\bm{x})\xi_j(t',\bm{x}')}=-2T\sigma'\delta_{ij}\partial_k^2\delta(t-t')\delta^{(3)}(\bm{x}-\bm{x}').
\end{split}
\end{equation}
$n_1(t,\bm{x})$ and $n_2(t,\bm{x})$ couple with each other through their streaming terms.
The equation of motion for $n_1(t,\bm{x})$ reads
\begin{equation}
\begin{split}
\left(\partial_0+ \sigma'\rho'\left(\partial_i^2\right)^2 \right)^2n_1(t,\bm{x}) = - n_3^2\rho'^2 \left(\partial_i^2\right)^2 n_1(t,\bm{x}).
\end{split}
\end{equation}
The dispersion relation is obtained as
\begin{equation}
\begin{split}
\omega  =  \pm v' k^2 - i\Gamma' k^4,
\end{split}
\label{eq:typeBdispersion}
\end{equation}
where $v'=|n_3|\rho'$ and $\Gamma'=\sigma'\rho'$.
Therefore, the type-B NG mode belongs to type-II NG modes as long as $v'\neq0$. 
We emphasize here that for both type-A and type-B NG modes,  the imaginary parts are smaller than the real parts at small $k$, so that the spectra become sharper as $k$ decreases.

In this section, we considered simple examples in classical models. In the following sections, we consider  general quantum systems at finite temperature, and show that  
the same dispersion relations~\eqref{eq:typeAdispersion} and~\eqref{eq:typeBdispersion} hold even in them.

\section{Spontaneous symmetry breaking and elastic variables}\label{sec:SSB}

In this section, we focus on how many independent-elastic variables appear when internal symmetries are spontaneously broken 
(For more general cases including spontaneous breaking of spacetime symmetries, see e.g, Ref.~\cite{Hayata:2013vfa}).
We see that the number of independent-elastic variables is equal to the number of broken symmetries, 
which is a generalization of Nambu-Goldstone theorem~\cite{Goldstone:1962es}
in vacuum to that for elastic variables at finite temperature.
(For the Nambu-Goldstone theorem at finite temperature, see Ref.~\cite{Strocchi:2005yk}.)

We assume that the Lagrangian is invariant under a symmetry group $G$, whose charges (generators)
 $\hQ_A$ are given by the integral of 
local charge densities $\hn_A(\bm{x})$,  
\begin{equation}
\begin{split}
\hQ_A=\int d^3x \,\hn_A(\bm{x}).
\end{split}
\end{equation}
We assume that $G$ breaks into a subgroup $H$. 
We employ indices with capital letters ($A,B,\cdots$) for charges of $G$, $\hQ_A$, with small letters ($a,b,\cdots$) for charges of $G/H$, $\hQ_a$,
and with Greek letters ($\alpha,\beta,\cdots$) for charges of $H$, $\hQ_\alpha$.
We use the hat symbol to indicate quantum operators to distinguish them from classical ones.

First, we consider the thermodynamic potential
\begin{equation}
\begin{split}
 W[J]=  -\frac{1}{\beta} \ln\tr\exp \left[-\beta \hK+ \beta\int d^3x \hPhi_l(\bm{x})J^l(\bm{x}) \right] ,
\end{split}
\end{equation}
where $\hK\equiv\hH-\mu\hN$. Here, $\hH$, $\hN$, $\beta=1/T$, $\mu$ are the Hamiltonian, the number operator, the inverse temperature, and the chemical potential, respectively.
$\hPhi_l(\bm{x})$ is a set of Hermitian local operators belonging to a linear representation, which may be either elementary or composite.
We choose that $\hPhi_l(\bm{x})$, at least, contains one order parameter for each broken symmetry.
The functional derivative of $-W[J]$ with respect to $J^l(\bm{x})$ gives the expectation value of $\hPhi_l(\bm{x})$:
\begin{equation}
\begin{split}
\average{\hPhi_l(\bm{x})}_J = -\frac{\delta W[J]}{\delta J^l(\bm{x})}.
\end{split}
\end{equation}
The subscript $J$ denotes the thermal average with the external field.
The second derivative of $-W[J]$ at $J=0$ gives the susceptibility,
\begin{equation}
\begin{split}
\chi_{lm}(\bm{x},\bm{x}')\equiv \left. -\frac{\delta^2  W[J]}{\delta J^l(\bm{x}) \delta J^m(\bm{x}')}\right|_{J=0} =
\int_0^\beta d\tau \average{e^{\tau \hK}\delta\hPhi_l(\bm{x})e^{-\tau \hK} \delta\hPhi_m(\bm{x}')}_{J=0}, \label{eq:susceptibilities}
\end{split}
\end{equation}
where $\delta\hPhi_l(\bm{x})\equiv\hPhi_l(\bm{x})-\average{\hPhi_l(\bm{x})}_{J=0}$.
The free energy is given by the Legendre transformation of $W[J]$:
\begin{equation}
\begin{split}
F[\Phi]=  W[J]- \int d^3x J^m(\bm{x})\frac{\delta W[J]}{\delta J^m(\bm{x})} . \label{eq:effectiveAction}
\end{split}
\end{equation}
The functional derivative of $ F[\Phi]$ with respect to $\Phi_l(\bm{x})$ is equal to $J^l(\bm{x})$, i.e.,
\begin{equation}
\begin{split}
\frac{\delta  F[\Phi]}{\delta\Phi_l(\bm{x})}=J^{l}(\bm{x}). \label{eq:dFdphi}
\end{split}
\end{equation}
At $J^l(\bm{x})=0$, this gives the stationary condition of the free energy.
The inverse of susceptibility is obtained by
\begin{equation}
\begin{split}
\chi^{lm}(\bm{x},\bm{x}') = \left.\frac{\delta^2F[\Phi]}{\delta \Phi_l(\bm{x})\delta\Phi_m(\bm{x}')}\right|_{\Phi=\average{\Phi}_{J=0}},
\label{eq:InverseSusceptibility}
\end{split}
\end{equation}
which satisfies
\begin{equation}
\begin{split}
\int d^3x''\chi_{lk}(\bm{x},\bm{x}'')\chi^{kn}(\bm{x}'',\bm{x}')=\delta_l^{~n}\delta^{(3)}(\bm{x}-\bm{x}').
\end{split}
\end{equation}

Next, let us consider the symmetry of the free energy.
When $\hQ_A$ commutes with $\hat{K}$, i.e., $[\hQ_A,\hat{K}]=0$, and if there is no quantum anomaly,
the free energy satisfies 
\begin{equation}
\begin{split}
\int d^3x'\frac{\delta F[\Phi] }{\delta\Phi_m(\bm{x}')} \OP_{Am}(\bm{x}';J)=0,
\label{eq:STIdentity}
\end{split}
\end{equation}
where $\OP_{Am}(\bm{x}';J)\equiv\average{[i\hQ_A,\hPhi_m(\bm{x}')]}_J$~\cite{WeinbergText}.
Taking the functional derivative of Eq.~\eqref{eq:STIdentity} with respect to  $\Phi_l(\bm{x})$,
we obtain
\begin{equation}
\begin{split}
\int d^3x' \frac{\delta^2 F[\Phi]}{\delta \Phi_l(\bm{x})\delta \Phi_m(\bm{x}')}\OP_{Am}(\bm{x}';J)
+\int d^3x' \frac{\delta F[\Phi]}{\delta \Phi_m(\bm{x}')} \frac{\delta}{\delta  \Phi_l(\bm{x})}\OP_{Am}(\bm{x}';J)
=0.
\label{eq:Fderivative}
\end{split}
\end{equation}
In the limit $J^l(\bm{x})\to0$, the second term in Eq.~\eqref{eq:Fderivative} vanishes from Eq.~\eqref{eq:dFdphi}, and it reads 
\begin{equation}
\begin{split}
\int d^3x' \frac{\delta^2 F[\Phi]}{\delta \Phi_l(\bm{x})\delta \Phi_m(\bm{x}')}\OP_{Am}(\bm{x}')
=0. \label{eq:FlattnessOfFreenergy}
\end{split}
\end{equation}
Here, we defined  $\OP_{Am}(\bm{x}')\equiv\lim_{J\to0}\average{[i\hQ_A, \hPhi_m(\bm{x}')]}_{J}$.
When the continuum symmetry is spontaneously broken, a nonvanishing $\OP_{al}(\bm{x})$ exists.
Furthermore, if the translational symmetry is not broken, $\OP_{al}(\bm{x})$ is constant.
From Eq.~\eqref{eq:InverseSusceptibility},  we can write Eq.~\eqref{eq:FlattnessOfFreenergy} as
\begin{equation}
\begin{split}
\chi^{lm}(\bm{k}=\bm{0})\OP_{am} =0.
\label{eq:zeroEigenvalue}
\end{split}
\end{equation}
Therefore, $\OP_{am}$ are eigenvectors of $\chi^{lm}(\bm{k}=\bm{0})$ with the zero eigenvalue, which 
represent the flat directions of free energy.
The number of independent-elastic variables is equal to the number of independent $\OP_{al}$. 
If there exists a linearly dependent vector for  an independent broken generator, i.e., $c^a \OP_{al}=0$ for nontrivial real $c^a$, 
the linear combination $c^a\hQ_a\neq0$ becomes an unbroken charge and thus this does not occur due to the definition of broken charges.
Therefore, the numberer of independent-elastic variables is equal to the number of broken generators.

At finite momentum, we can expand the inverse susceptibilities as
\begin{equation}
\begin{split}
\chi^{lm}(\bm{k}) = \rho^{lm}k^2  +\cdots  \geq0.
\end{split}
\end{equation}
Here, $\cdots$ denotes the higher-order terms in $k$. The eigenvalues of $\rho^{lm}$ are nonnegative because of the convexity of the free energy.
If $\rho^{lm}$ contains the zero eigenvalue, there appears a long-range correlation that causes the vanishing of order parameter for the three spatial dimensions. 
In general, when the susceptibility behaves like $\chi\sim k^{-\alpha}$ at small $k$, the infrared contributions in one-loop correction by thermal fluctuations of elastic variables to the order parameter is proportional to $\int_{\mu} d^{d}k/k^\alpha\sim \mu^{d-\alpha}$. Here, $\mu$ is the infrared cutoff, and $d$ denotes the spatial dimensions. When $d\leq \alpha$, the one-loop correction diverges at $\mu\to0$, and it leads to restoration of symmetry~\cite{PhysRevLett.17.1133,Coleman:1973ci}. 

Nonanalytic terms may appear in this expansion; however, 
in the following, we assume that at least the leading order of $\chi^{lm}(\bm{k})$ is quadratic in $k$, and the eigenvalues of the coefficient $\rho^{lm}$ are positive.

\section{Generalized Langevin equation}\label{sec:LangevinEquation}
In this paper, to derive the dispersion relations of NG modes, 
we employ the generalized Langevin equations for slow variables, which are formally obtained from Mori's projection operator method~\cite{Mori, Nordholm,Zwanzig,Rau:1995ea,Balucani2003409,Minami:2012hs}.
In this section, we briefly review the projection operator method.
Readers who are already familiar with it can skip to Sec.~\ref{sec:DispersionRelation}.

\subsection{Derivation}
We derive the generalized Langevin equations for a set of operators $\{\hA_n\}$.
The choice of operators are arbitrary. If one is interested in hydrodynamic behavior, one may choose the Hamiltonian (energy) density, momentum density, and all the other charge densities as a set of operators. We will choose elastic variables  and broken charge densities as $\{\hA_n\}$ in the next section.
In this subsection, we include coordinate index into the subscript $n$ in order to avoid complexity.

In order to define the projection operator, we, first, introduce an inner product satisfying  positive definiteness, $\innerProd{\hO_1,\hO_1}>0$ (if $\hO_1\neq0$), Hermite symmetry, $\innerProd{\hO_1,\hO_2 }=\innerProd{\hO_2,\hO_1}^* $, and linearity $\innerProd{a\hO_1+b\hO_2,\hO_3}=a\innerProd{\hO_1,\hO_3}+b\innerProd{\hO_2,\hO_3}$.
  The explicit form of the inner product will be given later.
Using the inner product, we define a metric as
\begin{equation}
g_{nm} \equiv \innerProd{{\hA}_n, {\hA}_m}.
\end{equation}
We also define $g^{ml}$ as the inverse of $g_{nm}$, i.e.,
$g_{nm}g^{ml}={\delta_{n}}^{l}$,
where Einstein's convention on repeated indices is understood. 
An operator with an upper index is defined as ${\hA}^{n} \equiv   g^{nm}{\hA}_{m}$.
By using the inner product, we
define the projection operator $\cP$ acting on a field $\hO$ as
\begin{equation}
\cP \hO \equiv  {\hA}_n \innerProd{\hO,{\hA}^n}.
\end{equation}
We also define $\cQ \equiv1-\cP$. These satisfy  $\cP^2=\cP$, $\cQ^2=\cQ$, and $\cQ\cP=\cP\cQ=0$. 
Using them, we can construct  the generalized Langevin equation,
\begin{equation}
\begin{split}
\partial_0 {\hA}_n(t)&= i{\varOmega_n}^m{\hA}_m(t)
-\int_0^\infty ds {\memory_n}^m(t-s) {\hA}_m(s)+\noise_n(t),
\label{eq:generalizedLangevin}
\end{split}
\end{equation}
from  the Heisenberg equation,
\begin{equation}
\begin{split}
\partial_0 {\hA}_n(t)=i[\hH,\hA_n(t)]\equiv  i\Lv \hA_n(t) \label{eq:heisenberg},
\end{split}
\end{equation}
where the Liouville operator $\Lv$ is introduced. 
The first, second, and third terms in Eq.~\eqref{eq:generalizedLangevin} are  the streaming, dissipation, and noise terms, respectively, where the frequency matrix $i{\varOmega_{n}}^m$, the memory function ${\memory_n}^m(t-s)$, and the noise operator $\noise_n(t)$ are given explicitly as
\begin{align}
i{\varOmega_{n}}^m&\equiv \innerProd{i\Lv \hA_n,\hA^m}, \label{eq:streaming}\\
{\memory_n}^m(t-s)&\equiv-\theta(t-s) \innerProd{i\Lv \noise_n(t-s), \hA^m},\\
\noise_n(t)&\equiv e^{\cQ i\Lv t}\cQ i\Lv \hA_n. \label{eq:Rn}
\end{align}
Here, $\theta(t)$ is Heaviside's step function.  The streaming term describes a time-reversible change, while the dissipation term does a time-irreversible change.
The memory function is the generalization of friction, and contains retarding effects.
Roughly speaking, the streaming and dissipation terms contribute to the real and imaginary parts of the dispersion relation, respectively.

Let us quickly derive the generalized Langevin equation~\eqref{eq:generalizedLangevin}, and functions~\eqref{eq:streaming} to \eqref{eq:Rn}.
We can formally solve Eq.~\eqref{eq:heisenberg} and find $\hA_n(t)=\exp(i\Lv t)\hA_n$.
Consider the Laplace transformations of $\exp({i \Lv t})$ and $\partial_0\exp({i \Lv t})=\exp({i \Lv t})i\Lv$, which can be written as
\begin{equation}
\frac{1}{z - i \Lv } = \frac{1}{z - i \Lv} \cP i\Lv \frac{1}{z -\cQ i\Lv }+\frac{1}{z -\cQ i\Lv }, \label{eq:oi1}\\
\end{equation}
and
\begin{equation}
\frac{1}{z-i\Lv}i\Lv=\frac{1}{z-i\Lv}\cP i\Lv+\frac{1}{z-i\Lv}\cQ i\Lv,  \label{eq:oi2}
\end{equation}
respectively.
From Eq.~\eqref{eq:oi1}, the second term in the right-hand side of Eq.~\eqref{eq:oi2} reads
\begin{equation}
\frac{1}{z - i \Lv }\cQ i\Lv = \frac{1}{z - i \Lv} \cP i\Lv \frac{1}{z -\cQ i\Lv }\cQ i\Lv+\frac{1}{z -\cQ i\Lv }\cQ i\Lv, \\
\end{equation} 
and thus, we obtain
\begin{equation}
\begin{split}
\frac{1}{z-i\Lv}i\Lv&=\frac{1}{z-i\Lv}\cP i\Lv+ \frac{1}{z - i \Lv} \cP i\Lv \frac{1}{z -\cQ i\Lv }\cQ i\Lv+ \frac{1}{z -\cQ i\Lv }\cQ i\Lv .
\label{eq:oi3}
\end{split}
\end{equation}
The inverse Laplace transformation of Eq.~\eqref{eq:oi3} leads to 
\begin{equation}
\partial_0 e^{i \Lv t}=e^{i \Lv t} \cP i \Lv
+\int^t_0ds e^{i\Lv s} \cP i \Lv  e^{ \cQ i \Lv (t-s)} \cQ i \Lv
+e^{ \cQ i \Lv t} \cQ i \Lv. \label{eq:operatoridentity}
\end{equation}
Multiplying ${\hA}_n$ by Eq.~\eqref{eq:operatoridentity}, 
we obtain Eq.~\eqref{eq:generalizedLangevin}~\cite{Mori}.
Remark here that Eq.~\eqref{eq:generalizedLangevin} is an operator identity for the solution of the Heisenberg equation,
and is satisfied for an arbitrary operator set. 
If one choose the inner product satisfying $\innerProd{i\Lv\hO_1,\hO_2}=-\innerProd{\hO_1,i\Lv\hO_2}$,
the memory function satisfies the fluctuation-dissipation theorem,
\begin{equation}
\begin{split}
{\memory_n}^m(t-s)&=\theta(t-s) \innerProd{ \noise_n(t-s), \noise^m},
\label{eq:Memory2}
\end{split}
\end{equation}
where we used the orthogonality of the noise operator, $\innerProd{\noise_n(t),{\hA}_m}=0$.

Taking the inner product of Eq.~\eqref{eq:generalizedLangevin} with $\hA^m$, 
we find that the Kubo response function~\cite{Kubo:1957mj},
 ${G_n}^m(t)=\innerProd{{\hA}_n(t),{\hA}^m}$, satisfies
\begin{equation}
\begin{split}
\partial_0 {G_n}^m(t)&= i{\varOmega_n}^l{G_l}^m(t)-\int_0^\infty ds {\memory_n}^l(t-s) {G_l}^m(s) ,
\label{eq:moriProjectionPropagator}
\end{split}
\end{equation}
where we also used $\innerProd{\noise_n(t),{\hA}_m}=0$.
Equation~\eqref{eq:moriProjectionPropagator} can be also obtained from a different method called the memory function method~\cite{mazenko}.

In the Laplace space, Eq.~\eqref{eq:moriProjectionPropagator} reads
\begin{equation}
\begin{split}
\big(z{\delta_{n}}^l-i{\varOmega_n}^l+{\memory_n}^l(z) \big){G_l}^m(z)
 ={\delta_n}^m.
\label{eq:GreenFunctionIdentity}
\end{split}
\end{equation}
The excitation modes are obtained by the roots of 
\begin{equation}
\begin{split}
\det\big(z{\delta_{n}}^m-i{\varOmega_n}^m+{\memory_n}^m(z)\big)=0  \label{eq:detG}
\end{split}
\end{equation}
 in the complex $z$-plane.
Therefore, the noise term needs not to be taken into account  when the dispersion relations of excitation modes are discussed.
Note that not only low-energy excitations  but also all higher-excitation modes coupled to $\{\hA_n\}$ are roots of the determinant.
Only if one chooses slow variables as $\{\hA_n\}$, one can apply the low-energy expansion to Eq.~\eqref{eq:GreenFunctionIdentity} or~\eqref{eq:detG} and calculates the dispersion relations of low-energy excitations,
which will be discussed in Sec.~\ref{sec:LowEnergyExpansion}.

At finite temperature and/or density, it is useful to employ the Kubo-Mori-Bogoliubov inner product 
\begin{equation}
\innerProd{\hO_1,\hO_2}  \equiv \frac{1}{\beta}\int_0^\beta d\tau \average{e^{\tau \hK}\hO_1e^{-\tau \hK} \hO_2^\dag}.
\label{eq:innerProduct}
\end{equation} 
Using this inner product, the frequency matrix reads
\begin{equation}
\begin{split}
i{\varOmega_{n}}^m&=i\frac{1}{\beta}\int_0^\beta d\tau \average{e^{\tau\hK}[\hH,\hA_n]e^{-\tau\hK}\hA^{m\dag}}\\
&=i\frac{1}{\beta}\int_0^\beta d\tau \partial_\tau \average{e^{\tau\hK}\hA_ne^{-\tau\hK}\hA^{m\dag}}
+i\mu\innerProd{[\hN,\hA_n],\hA^m}\\
&=-i\frac{1}{\beta} \average{[\hA_n,\hA^{m\dag}]}
+i\mu q_{n}^{~m},
\end{split}\label{eq:streaming2}
\end{equation}
where we assumed $[\hN,\hA_n]=q_n^{~m}\hA_m$.
Therefore, the streaming term can be written as
\begin{equation}
\begin{split}
i{\varOmega_n}^m{\hA}_m(t)=\left.\{A_n, F(A)\}_P\right|_{A_n=\hA_n(t)} + i\mu q_{n}^{~m}\hA_m(t)
\label{eq:StreamingChemical}
\end{split}
\end{equation}
with the quadratic free energy $F=T {A^{m\dag}}A_m$ and the Poisson bracket $\{A_n, A^\dag_m\}_P\equiv -i\average{[\hA_n,\hA_m^\dag]}$.
In this formulation, the nonvanishing expectation values of the commutation relations of $\hA_{n}$ give the canonical relations.

When one chooses fluctuations, $\hPhi_n-\average{\hPhi_n}$, as $\hA_n$, 
the inner product of $\hA_n$'s is related to the susceptibility in Eq.~\eqref{eq:susceptibilities} by $\beta\innerProd{\hA_n,\hA_m} = \chi_{nm}$.
${G_{n}}^m(t)$ determines the time evolution of a nonequilibrium state with the density operator,
\begin{equation}
\begin{split}
\hrho_\text{init}= \hrho_\text{eq}+  \frac{1}{\beta}\int_0^\beta d\tau \hrho_\text{eq} e^{\tau\hK}\hA^{n\dag}e^{-\tau\hK} A^{(0)}_n,
\end{split}
\label{eq:initial}
\end{equation}
where $\hrho_\text{eq}\equiv \exp(-\beta \hK)/\tr \exp(-\beta \hK)$, and
$A^{(0)}_n=\langle \hA_{n}\rangle_\text{init} \equiv\tr \hrho_\text{init}\hA_{n}$.
The expectation value at $t>0$ is expressed as
\begin{equation}
\langle \hA_n(t)\rangle_\text{init} = {G_{n}}^m(t) A^{(0)}_m,
\end{equation}
where we used $\average{\hA_n}=0$.
Note that the Kubo response function ${G_{n}}^m(t)$ is different from the retarded Green function, whose poles are often used to define the dispersion relations of excitation modes. However, the locations of poles of ${G_{n}}^m(t)$ coincide with  those of the retarded Green function (see Appendix.~\ref{sec:RelationBetweenResponseFunctions}.)

\subsection{Low-energy expansion} \label{sec:LowEnergyExpansion}
We here perform the low-energy expansion. 
For this purpose, we work in Laplace-momentum space, in which the Langevin equation is expressed as
\begin{equation}
\begin{split}
\bigl[z\delta_n^{~m}-(i\Omega_{nl}(\bm{k})- \memory_{nl}(z,\bm{k}))g^{lm}(\bm{k})   \bigr]\hA_m(z,\bm{k}) = \hA_n(t=0,\bm{k})+\noise_n(z,\bm{k}),
\label{eq:Langevin3}
\end{split}
\end{equation}
where we explicitly revived the inverse metric $g^{nm}(\bm{k})$. 
We assume that $g^{nm}(\bm{k})$, $i\Omega_{nm}(\bm{k})$ and $\memory_{nm}(z,\bm{k})$ can be expanded around $z=0$ and $\bm{k}=\bm{0}$.
Then, $g^{nm}(\bm{k})$ and $i\Omega_{nm}(\bm{k})$ can be expanded in powers of $k$ as
\begin{align}
g^{nm}(\bm{k}) &= g^{(0)nm} +g^{(2)nm}k^2  +\cdots, \\
i\Omega_{nm}(\bm{k}) &= i\Omega^{(0)}_{nm}+ i\Omega^{(2)}_{nm}k^2 + \cdots,
\end{align}
where we assumed that $\hA_n(z,\bm{k})$ are scalar fields that have no spatial index. For vector or tensor fields, the linear term in $k^i$, $g^{(1)nm}$ and $i\Omega^{(1)}_{nm}$, may appear. 
In fact, this is the case for the hydrodynamic equations derived from the projection operator method~\cite{Minami:2012hs}, where the momentum density, $\hat{p}^i(t,\bm{x})$, has the spatial index.
As was discussed in the previous section, the (inverse) metric is related to the generalized susceptibility, and thus the one for the elastic variables vanishes at $\bm{k}=\bm{0}$. 

Here, we only consider the case for $[\hN, \hA_n(t,\bm{x})]=0$.  
It is easy to generalize to the case for $[\hN,\hA_n(t,\bm{x})]=q_n^{~m}\hA_m(t,\bm{x})$ by shifting $z\delta_n^{~m}$  to $z\delta_n^{~m}-i\mu q_n^{~m}$ in Eq.~\eqref{eq:Langevin3} 
because the chemical potential only shifts the frequency matrix $i\varOmega_n^{~m}$ to $i\varOmega_n^{~m}+i\mu q_n^{~m}$ from Eq.~\eqref{eq:streaming2}.
$i\Omega^{(0)}_{nm}$ reads 
\begin{equation}
\begin{split}
i\Omega^{(0)}_{nm}= -\frac{i}{\beta}\int d^3x \average{[\hA_n(\bm{x}),\hA^\dag_m(\bm{0})]}.
\end{split}
\end{equation}
If $\hA_n(\bm{x})$ and $\hA_m(\bm{x}')$ are a charge density $\hn_a(\bm{x})$ and an elastic variable $\hpi_i(\bm{x}')$, $i\Omega^{(0)}_{nm}$ becomes
\begin{equation}
\begin{split}
i\Omega^{(0)}_{n_a\pi_i}
= -\frac{1}{\beta}\average{[i\hQ_a,\hpi_i(\bm{0})]}\neq 0.
\end{split}
\end{equation}
This is nothing but the condition of spontaneous breaking, Eq.~\eqref{eq:SSB}.
Therefore, 
when the symmetry is spontaneously broken, the charge density and the local operator become canonically conjugate with each other
because $i\Omega^{(0)}_{n_a\pi_i}$ corresponds to the poisson bracket in the Langevin equation as in Eq.~\eqref{eq:StreamingChemical}.

Next, we consider the memory function.  
We perform the expansion with respect to $z$:
\begin{equation}
\memory_{nm}(z,\bm{k}) = i\delta \varOmega_{nm}(\bm{k})+L_{nm}(\bm{k})+z\delta Z_{nm}(\bm{k})+ \mathcal{O}(z^2),
\end{equation}
where we decomposed the leading part into $i\delta \varOmega_{nm}(\bm{k})=(\memory_{nm}(0,\bm{k})-\memory^*_{mn}(0,\bm{k}))/2$ and $L_{nm}(\bm{k})=(\memory_{nm}(0,\bm{k})+\memory^*_{mn}(0,\bm{k}))/2$.
$i\delta \varOmega_{nm}(\bm{k})$ can be renormalized into the frequency matrix, $i\bar{\varOmega}_{nm}(\bm{k})\equiv i\varOmega_{nm}(\bm{k})-i\delta{\varOmega}_{nm}(\bm{k})$, and can be expanded at small $k$. $L_{nm}(\bm{k})$ contributes to dissipation.
$\delta Z_{nm}(\bm{k})$ gives the correction of the time derivative term.
We assumed that ${\memory_{nm}}(z,\bm{k}) $ does not contain terms proportional to the inverse power of $z$ corresponding to additional zero modes. If $\memory_{nm}(z,\bm{k}) $ contains such a zero mode, we need to treat it as the independent slow variable, and project  the mode out from the memory function.

$L_{nm}(\bm{k})$ is expanded as
\begin{equation}
\begin{split}
L_{nm}(\bm{k}) = L^{(0)}_{nm}+L^{(2)}_{nm}k^2  + \cdots.
\end{split}
\end{equation}
Since the memory function~\eqref{eq:Memory2} is the correlation of noises, and the noise operator contains the time derivative of the field, $\partial_0\hA_n= i\Lv\hA_n$,
if $\hA_n$ is taken to be a charge density, it vanishes at the low momentum limit. This is due to the current conservation, $\Lv\hn_a(z,\bm{k})=-k^i \hat{j}^i_a(z,\bm{k})$ in momentum space.
The memory function for charge densities $K_{n_an_b}(z,\bm{k})$ is expressed as
\begin{equation}
\begin{split}
K_{n_an_b}(z,\bm{k})= k^2 \tilde{K}_{n_an_b}(z,\bm{k})
\end{split}
\end{equation}
with 
\begin{equation}
\begin{split}
\tilde{K}_{n_an_b}(z,\bm{k}) \equiv \frac{k^ik^j}{k^2}\int_0^\infty dt e^{-zt} \int d^3x e^{-i\bm{k}\cdot\bm{x}} \innerProd{ e^{\cQ i\Lv t}\cQ \hat{j}_{n_a}^i(t,\bm{x}),\hat{j}_{n_b}^j(0,\bm{0})}.
\end{split}
\end{equation}
Therefore, $K_{n_an_b}(z,\bm{k})$ is at least of order $k^2$.
Similarly, the memory function between a charge density $\hn_a$ and a local operator $\hphi_i$,
$K_{n_a\phi_i}(z,\bm{k})$, is also of order $k^2$. 
Thus,  $\delta\varOmega_{nm}^{(0)}$, $L_{nm}^{(0)}$, and $\delta Z_{nm}^{(0)}$ vanish when $\hA_n$ or $\hA_m$ is a charge density.
The nonvanishing $\delta\varOmega_{nm}^{(0)}$, $L_{nm}^{(0)}$, and $\delta Z_{nm}^{(0)}$ are possible only for the memory function of $\phi$'s,
i.e., $K_{\phi_i\phi_j}(z,\bm{k})$.  

The left-hand side in Eq.~\eqref{eq:Langevin3} is expanded up to the order of $z$ and $k^2$ as
\begin{equation}
\begin{split}
&z\delta_n^{~m}-(i\Omega_{nl}(\bm{k})- \memory_{nl}(z,\bm{k}))g^{lm}(\bm{k}) \\
&\quad=z(\delta_n^{~m}+\delta Z_{nl}g^{(0)lm})
 -i\bar{\Omega}^{(0)}_{nl}g^{(0)lm}+L^{(0)}_{nl}g^{(0)lm}\\
&\qquad+k^2\Bigl(
 -i\bar{\Omega}^{(2)}_{nl}g^{(0)lm}+L^{(2)}_{nl}g^{(0)lm}
 -i\bar{\Omega}^{(0)}_{nl}g^{(2)lm}+L^{(0)}_{nl}g^{(2)lm}
 \Bigr).
\end{split}
\end{equation}
The roots of this matrix in the complex $z$ plane give the dispersion relations of low-energy excitation modes.

\section{Nambu-Goldstone modes and their dispersion relations}\label{sec:DispersionRelation}
In this section, we discuss type-A and type-B NG modes, their dispersion relations,  the gap formula, and the mixing with the hydrodynamic mode at finite temperature and/or density. 
We also discuss the existence of gapped partners of type-B NG modes, when the expectation values of a charge density and a local operator that break the same symmetry
 coexist.

\subsection{Classification of broken charges}
We classify the broken charges into two types: type-A and type-B, which correspond to type-A and type-B in the classification of NG modes, respectively.
We consider the situation in which a symmetry group $G$ is spontaneously broken into a subgroup $H$ as in the case of Sec.~\ref{sec:SSB}.
We write $T_A$, $S_{\alpha}$, and $X_a$ as the generators of $G$, $H$, and $G/H$, respectively.
The spontaneous symmetry breaking is characterized by the nonvanishing expectation value 
of commutation relation between the charge and a local operator. 
There are two possibilities for the local operator: Either it is a charge density itself or it is not.
As was discussed with examples in Sec.~\ref{sec:Examples}, the dispersion relations of NG modes are different in these cases. 
To study these cases, we introduce the following operator:
\begin{equation}
\begin{split}
\hPhi(\bm{x})\equiv \begin{bmatrix}
\hphi_i(\bm{x})\\
\hn_A(\bm{x})\\
\end{bmatrix},
\end{split}
\end{equation}
where  $\hphi_i({\bm{x}})$ are local operators that have different quantum numbers from charge densities and belong to a real representation R.
$\hn_A(\bm{x})$ are all charge densities of $G$.
This operator transforms under $\hQ_B$ as 
\begin{equation}
\begin{split}
[i\hQ_B, \hPhi(\bm{x})]= iT_B\hPhi(\bm{x})=  \begin{bmatrix}
[iT^{\text{R}}_B]_i^{~j}\hphi_j(\bm{x})\\
-[iT^{\text{adj}}_B]_{A}^{~C}\hn_C(\bm{x})\\
\end{bmatrix},
\end{split}
\end{equation}
where $[T^{\text{adj}}_A]_B^{~C}=if_{BA}^{~~C}$ with structure constant $f_{AB}^{~~C}$.
Since we chose a real representation, $[iT^{\text{R}}_B]_i^{~j}$ and $-[iT^{\text{adj}}_B]_{A}^{~C}$ are real and antisymmetric.
The expectation value,
\begin{equation}
\begin{split}
\Phi_0\equiv \average{\hPhi(\bm{x})}= 
\begin{bmatrix}
\phi^0_{i}\\
n^0_{A}\\
\end{bmatrix},
\end{split}
\end{equation}
 is invariant under the unbroken symmetry, i.e., $S_\alpha\Phi_0=0$.
The expectation values of commutation relations between the broken charges $\hQ_a$ and $\hPhi(\bm{x})$ are written as
\begin{equation}
\begin{split}
\average{[i\hQ_a,\hPhi(\bm{x})]}= iX_a\Phi_0=
  \begin{bmatrix}
[iX^{\text{R}}_a]_i^{~j}\phi^0_j\\
-[iX^{\text{adj}}_a]_{A}^{~C}n^0_C\\
\end{bmatrix}.
\end{split}
\end{equation}
As was discussed in Sec.~\ref{sec:SSB}, $\average{[i\hQ_a,\hPhi(\bm{x})]}$ are the eigenvectors of the second derivative of the free energy with the zero eigenvalue, and are
the bases of $\NBS$-dimensional real-vector space.

We classify the charges into two types: $\hQ^\typeA_a$ and $\hQ^{\typeB}_{\bar{a}}$,  which are defined such that 
the bases have the form,
\begin{align}
\text{Type A}:&\ \Phi_{(a)}^\typeA \equiv 
\average{[i\hQ^\typeA_a,\hPhi(\bm{x})]}=
 \begin{bmatrix}
h^\typeA_{ai}\\
0\\
\end{bmatrix}, \label{eq:TypeAfield}\\
\text{Type B}:&\ \Phi_{(\bar{a})}^{\typeB}  \equiv 
\average{[i\hQ^{\typeB}_{\bar{a}},\hPhi(\bm{x})]}
=  \begin{bmatrix}
h^\typeB_{\bar{a}i}\\
W_{\bar{a}\bar{b}}^{\typeB}\\ 
\end{bmatrix},  \label{eq:classification}
\end{align} 
and $\Phi_{(a)}^\typeA$ are orthogonal to $\Phi_{(\bar{a})}^\typeB$, $(\Phi_{(a)}^{\typeA })^T\Phi_{(\bar{b})}^\typeB=0$.
This classification enables us to separate the type-A and type-B sectors in the equations of motion.

This decomposition can be done as follows:
First, we consider the expectation values of commutation relations between broken charges and their charge densities,
\begin{equation}
\average{[i\hQ_a,\hn_b(\bm{x})]}\equiv  W_{ab}= -[iX_a]_b^{~c}n^0_{c}.
\label{eq:commutationRelationOfCharges}
\end{equation}
Here, $W_{ab}$ is a real antisymmetric matrix, which can be decomposed into
\begin{equation}
\begin{split}
{O_a}^c W_{cd}{O^d}_b =
\begin{bmatrix}
0 & 0\\
0& W^\text{B} 
\end{bmatrix}_{ab},
\end{split}
\end{equation}
with 
\begin{equation}
W^{\text{B}}=
\begin{bmatrix}
0 & \Lambda\\
- \Lambda& 0 
\end{bmatrix}_{ab},
\end{equation}
where ${O_a}^b$ is the orthogonal matrix, ${O^a}_c{O_c}^b=\delta^{ab}$, and $\Lambda=\mathrm{diag}(\lambda_1,\lambda_2,\cdots,\lambda_{N_\typeB})$
with $N_\typeB\equiv\rank W^\typeB/2=\rank\average{[i\hQ_a,\hQ_b]}/2$ being the number of pairs of type-B charges.
Using this matrix, we define type-A and type-B charge densities as
\begin{align}
\hn^{\typeA}_a(\bm{x}) &\equiv {O_a}^b \hn_b(\bm{x}),\\
\hn^{\typeB}_{\bar{a}}(\bm{x}) &\equiv {O_{\bar{a}+N_\typeA}}^b \hn_b(\bm{x}),
\end{align}
where $N_\typeA\equiv \NBS-\rank W$ is the number of type-A charges.
In the following, we distinguish type-A charge densities and those of type-B by bar indices.

In this decomposition, the expectation values of commutation relations between $\hQ^\typeA_a$ and $\hPhi_i({\bm{x}})$ have the form of Eq.~\eqref{eq:TypeAfield}.
In general, $\Phi_{(\bar{a})}^\typeB\equiv\average{[i\hQ^\typeB_{\bar{a}},\hPhi({\bm{x}})]}$ is not orthogonal to  $\Phi^{\typeA}_{(a)}$, i.e., 
the inner product, $(\Phi_{(a)}^\typeA)^T \Phi_{(\bar{a})}^\typeB = \sum_{i}\OP^\typeA_{ai}\OP^\typeB_{\bar{a}i} \equiv \eta_{a\bar{a}}$, 
does not vanish.
However, it can be taken to be zero by choosing the linear combination of charges, $\hQ'^\typeB_{\bar{a}}=\hQ^\typeB_{\bar{a}} - c_{~\bar{a}}^{b}\hQ_b^\typeA$, as the new type-B charge instead of $\hQ^\typeB_{\bar{a}}$.
Note that by this redefinition of type-B charges, the expectation values of commutation relations between broken charges and their charge densities do not change,
i.e., $\average{[i\hQ'^\typeB_{\bar{a}}, \hn'^\typeB_{\bar{b}}(\bm{x})]} = \average{[i\hQ^\typeB_{\bar{a}}, \hn^\typeB_{\bar{b}}(\bm{x})]}$
because $\average{[i\hQ^\typeB_{\bar{a}}, \hn^\typeA_{b}(\bm{x})]}=\average{[i\hQ^\typeA_{{a}}, \hn^\typeA_{b}(\bm{x})]}=0$.
Concretely, we introduce $\eta_{ab}\equiv
(\Phi_{(a)}^\typeA)^T \Phi_{(b)}^\typeA=
\sum_{i} \OP^\typeA_{ai} \OP^\typeA_{bi}$, which is a $N_\typeA\times N_\typeA$ regular matrix because of linear independence of $\Phi_{(a)}^\typeA$.
When we choose $c_{~\bar{a}}^{b}=\eta^{bc}\eta_{c\bar{a}}$,
$\Phi_{(\bar{a})}'^\typeB\equiv\average{[i\hQ'^{\typeB}_{\bar{a}},\hPhi({\bm{x}})]}=\Phi_{(\bar{a})}^\typeB-\Phi_{(b)}^\typeA \eta^{bc}\eta_{c\bar{a}}$ is orthogonal to $\Phi_{(a)}^\typeA$, where $\eta^{ab}$ is the inverse matrix of $\eta_{ab}$.
In the following, we omit the prime symbol from $\hQ'^\typeB_{\bar{a}}$. 

The order parameters $\Phi^{\typeA}_{(a)}$ and $\Phi^{\typeB}_{(\bar{a})}$ linearly transform under $H$ as 
$S_{\alpha}X_{b}\Phi_0= if_{\alpha b}^{~~c}X_c\Phi_0= if_{\alpha b}^{~~c}\Phi^\typeA_{(c)}+if_{\alpha b}^{~~\bar{c}}\Phi^\typeB_{(\bar{c})}$ with the structure constant $[S_{\alpha}, X_b]=if_{\alpha b}^{~~c}X_c$, where we used $S_{\alpha}\Phi_0=0$.
In general, $X_{b}\Phi_0$ are reducible; in particular, $\Phi_{(a)}^\typeA$ and $\Phi_{(\bar{a})}^{\typeB}$  belong to different representations, respectively.
For  $\Phi_{(b)}^\typeA$, 
since $X^\typeA_{b}n^0=0$, $S_{\alpha}X^A_{b}n^0= if_{\alpha b}^{~~c}X^\typeA_cn^0+if_{\alpha b}^{~~\bar{c}}X^\typeB_{\bar{c}}n^0=0$.
This implies $if_{\alpha b}^{~~\bar{c}}X^\typeB_{\bar{c}}n^0=0$, so that $if_{\alpha b}^{~~\bar{c}}=0$ because $X^\typeB_{\bar{c}}n^0$ are linearly independent.
Thus, $S_\alpha\Phi_{(b)}^\typeA=if_{\alpha b}^{~~c}\Phi_{(c)}^\typeA$ is satisfied.
We can also see that $\Phi^{\typeB}_{(\bar{a})}$ transforms under $H$ as $S_{\alpha}\Phi^{\typeB}_{(\bar{b})}= if_{\alpha \bar{b}}^{~~\bar{c}}\Phi^{\typeB}_{(\bar{c})}$ 
because $S_{\alpha}\Phi^{\typeB}_{(\bar{b})}$ is orthogonal to $\Phi^{\typeA}_{(a)}$:
$(\Phi^\typeA_{(a)})^T S_{\alpha}\Phi^\typeB_{\bar{b}}
=-(S_{\alpha}\Phi^\typeA_{(a)})^T\Phi^\typeB_{\bar{b}}
=-if_{\alpha a}^{~~c}(\Phi^\typeA_{(c)})^T\Phi^\typeB_{\bar{b}}
=0$.

\subsection{Equations of motion for NG modes and their dispersion relations}
In this subsection, we derive the dispersion relations of NG modes. 
For this purpose, we employ the generalized Langevin equations and the low-energy expansion discussed in Sec.~\ref{sec:LangevinEquation}.
We assume that the type-A and type-B sectors are separated by an unbroken symmetry.
In order to define the NG fields, we introduce the projection operator $\mathcal{P}^{\bar{a}}_{~\bar{c}}$ mapping to the kernel of $\eta_{\bar{a}\bar{b}}\equiv\sum_i h^\typeB_{\bar{a}i}h^\typeB_{\bar{b}i}$, i.e., $\eta_{\bar{a}\bar{b}}\mathcal{P}^{\bar{b}}_{~\bar{c}}=0$.
We also define $\eta^{\bar{a}\bar{b}}$ such that $\eta^{\bar{a}\bar{b}}\eta_{\bar{b}\bar{c}}=\mathcal{Q}^{\bar{a}}_{~\bar{c}}$, where 
$\mathcal{Q}^{\bar{a}}_{~\bar{c}}\equiv\delta^{\bar{a}}_{~\bar{c}}-\mathcal{P}^{\bar{a}}_{~\bar{c}}$.
These satisfy $\mathcal{P}^{\bar{a}}_{~\bar{c}}\mathcal{P}^{\bar{c}}_{~\bar{b}}=\mathcal{P}^{\bar{a}}_{~\bar{b}}$,
$\mathcal{Q}^{\bar{a}}_{~\bar{c}}\mathcal{Q}^{\bar{c}}_{~\bar{b}}=\mathcal{Q}^{\bar{a}}_{~\bar{b}}$, and $\mathcal{P}^{\bar{a}}_{~\bar{c}}\mathcal{Q}^{\bar{c}}_{~\bar{b}}=\mathcal{Q}^{\bar{a}}_{~\bar{c}}\mathcal{P}^{\bar{c}}_{~\bar{b}}=0$.
We define type-A and type-B NG fields by
\begin{align}
\hphi^{\typeA a}(t,\bm{x}) &\equiv \sum_i \eta^{ab}h^\typeA_{bi}\hphi_i(t,\bm{x}),\\
\hphi^{\typeB \bar{a}}(t,\bm{x}) &\equiv \sum_i \eta^{\bar{a}\bar{b}}h^\typeB_{\bar{b}i}\hphi_i(t,\bm{x}).
\end{align}
We choose $\delta\hn_a^\typeA(t,\bm{x})\equiv \hn_a^\typeA(t,\bm{x})-\average{\hn_a^\typeA(t,\bm{x})}$, 
$\deltaB\hn_{\bar{a}}^\typeB(t,\bm{x})\equiv \hn_{\bar{a}}^\typeB(t,\bm{x})-\average{\hn_{\bar{a}}^\typeB(t,\bm{x})}$,
$\hphi^{\typeA {a}}(t,\bm{x})$, and $\hphi^{\typeB {\bar{a}}}(t,\bm{x})$ as slow variables.
(Since the generators of a real representation $[iT^{\text{R}}_A]_i^{~j}$ are antisymmetric,
$\sum_ih^\typeA_{bi}\phi_i^0=\sum_ih^\typeB_{bi}\phi_i^0=0$.
They lead to $\average{\hphi^{\typeA a}(t,\bm{x}) }=\average{\hphi^{\typeB \bar{a}}(t,\bm{x})}=0$, so that $\delta$ is not necessary for these operators.)
We consider the case that the chemical potential couples to a $U(1)$ charge $\hN$ that commutes with other charges $\hQ_A$, i.e., $[\hN,\hQ_A]=0$.
In this case, we can apply the result obtained in Sec.~\ref{sec:SSB}.

The total number of dynamical degrees of freedom is $\NBS+\rank\average{[i\hQ_a,\hphi_i(\bm{x})]}$.
We will find that the number of type-A and type-B NG modes are equal to  $N_\typeA=\NBS-\rank\average{[i\hQ_a,\hQ_b]}$, and $N_\typeB=\rank\average{[i\hQ_a,\hQ_b]}/2$, respectively. Other degrees of freedom become gapped modes, whose number is given by $(\rank\average{[i\hQ_a,\hphi_i(\bm{x})]}-N_\typeA)/2$.
The gap becomes small, when $n^0$ is much smaller than the typical scale of the system such as the energy of non-NG mode, and thus they play roles of  low-energy degrees of freedom.

\subsubsection{Dispersion relations of type-A NG modes}
Let us start with the type-A NG modes. Pions in QCD and the phonon in the superfluid phase are examples of type-A NG modes.
To avoid complicated indices, we simply omit index `A', and use the matrix notation.
We first evaluate the frequency matrix. In the leading order of derivative expansion, $i\Omega^{(0)}_{ n \phi}$ is calculated  as
\begin{equation}
\begin{split}
i\Omega^{(0)}_{ n \phi}&= -\frac{1}{\beta}\average{[i\delta \hQ,\hphi(\bm{x})]}
= -\frac{1}{\beta} .
\end{split}
\end{equation}
From the definition of type-A charges, $i\Omega^{(0)}_{ n  n}$ vanishes.
In contrast,  $i\Omega^{(0)}_{\phi \phi}$ cannot be determined from the symmetry breaking.
If one chooses the operators such as $[\hphi_i(\bm{x}), \hphi_j(\bm{x}')]=0$ and $[\hN,\hphi_i(\bm{x})]=0$, $i\Omega^{(0)}_{\phi \phi}$ vanishes.

The flatness of  the free energy implies that  the inverse susceptibility behaves like $\chi^{\phi\phi}(\bm{k})=\stiffness k^2+\cdots=\beta g^{\phi\phi}(\bm{k})$, and thus,
$g^{(0)\phi\phi}=0$ and $g^{(2)\phi\phi} = \beta \stiffness$, where $\stiffness$ is the stiffness matrix, whose eigenvalues are positive unless the parameters are fine-tuned.
On the other hand, the inverse charge susceptibility is generally nonzero at $\bm{k}=\bm{0}$: $g^{(0)n n} =\beta\chi^{n n}\neq0$.
Note that the cross terms, $g^{\phi n}$ and $g^{n\phi},$ do not appear because $\hphi(\bm{x})$ are chosen as the eigenvectors of the inverse susceptibilities.
The wave-function correction $\delta Z^{(0)}_{\phi\phi}$ may be nonzero, while $\delta Z^{(0)}_{n \phi}=\delta Z^{(0)}_{n n}=0$ because of the charge conservation as discussed in the previous section.
However,  $\delta Z^{(0)}_{\phi\phi}$ does not contribute to the equations of motion in the leading order because $\delta Z_{\phi}^{(0)\phi}=\delta Z^{(0)}_{\phi\phi}g^{(0)\phi\phi}=0$ and $\delta Z_{\phi}^{(0)n}=\delta Z^{(0)}_{\phi\phi}g^{(0)\phi n}=0$.

The dispersion relations can be obtained from the roots of Eq.~\eqref{eq:detG}.
However, instead of solving this, we solve the corresponding equations of motion: 
\begin{equation}
\begin{split}
\partial_0
\begin{bmatrix}
\hphi(t,\bm{k})\\
\delta\hn(t,\bm{k})
\end{bmatrix}
&=
\begin{bmatrix}
i\bar{\Omega}^{(0)}_{\phi \phi }& 1\\
-1&0\\
\end{bmatrix}
\begin{bmatrix}
\stiffness k^2&0 \\
0&\chi^{n n}
\end{bmatrix}
\begin{bmatrix}
\hphi(t,\bm{k})\\
\delta\hn(t,\bm{k})
\end{bmatrix}
\\
&=
\begin{bmatrix}
i\bar{\Omega}^{(0)}_{\phi\phi}\stiffness k^2& \chi^{nn} \\
-\stiffness k^2& 0
\end{bmatrix}
\begin{bmatrix}
\hphi(t,\bm{k})\\
\delta\hn(t,\bm{k})
\end{bmatrix},
\end{split}
\label{eq:EOMtypeA}
\end{equation}
where we neglected the dissipation terms, which will be taken into account later.
The equation of motion for $\hphi(\omega,\bm{k})$ reads
\begin{equation}
\begin{split}
\bigl[-\omega^2- \bar{\Omega}^{(0)}_{\phi\phi}\stiffness \omega k^2+v^2k^2\bigr]\hphi(\omega,\bm{k})=0,
\end{split}
\end{equation}
where $v^2\equiv \chi^{n n} \stiffness$ is the velocity matrix.
The second term, $ \bar{\Omega}^{(0)}_{\phi\phi}\stiffness \omega k^2$ is of order $k^2\omega$, so that it is negligible to derive the dispersion relations in the leading order of small $k$.
The dispersion relations for type-A NG modes are given as  $\omega =\pm v_i k$, where $v_i$ are the eigenvalues of $\sqrt{v^2}$.
Note that the eigenvalues of $v^2$ are positive because our equations of motion are equivalent to those in a Hamiltonian system with a positive semidefinite Hessian matrix, when dissipation terms are neglected. Therefore, we can take the square root of $v^2$.

The Poisson bracket and the effective free-energy that reproduce these equations of motion are given as
\begin{equation}
\begin{split}
\{\phi^b(\bm{x}), \delta n_a(\bm{x}')\}_P = \delta_{~a}^{b}\delta(\bm{x}-\bm{x}'),
\end{split}
\end{equation}
and 
\begin{equation}
\begin{split}
F[\phi,n] = \int d^3x\Bigl( \frac{1}{2}\chi^{n_an_b} \delta n_a(\bm{x})\delta n_b(\bm{x}) +\frac{1}{2}\stiffness_{ab} \partial_i\phi^a(\bm{x}) \partial_i\phi^b(\bm{x})
\Bigr),
\end{split}
\end{equation}
respectively. 
These correspond to the canonical relation~\eqref{eq:PoissonBracket1} and the free energy~\eqref{eq:ElasticFreenergy},
discussed in Sec.~\ref{sec:Examples}, respectively.

Next, taking into account the dissipation effects, we obtain the following equations of motion:
\begin{equation}
\begin{split}
\partial_0
\begin{bmatrix}
\hphi(t,\bm{k})\\
\delta\hn(t,\bm{k})
\end{bmatrix}
&=
\begin{bmatrix}
i\bar{\Omega}^{(0)}_{\phi \phi }-L^{(0)}_{\phi \phi }& 1-L^{(2)}_{\phi  n }k^2\\
-1-L^{(2)}_{n\phi }k^2& -L^{(2)}_{nn}k^2\\
\end{bmatrix}
\begin{bmatrix}
\stiffness k^2&0 \\
0&\chi^{nn}
\end{bmatrix}
\begin{bmatrix}
\hphi(t,\bm{k})\\
\delta\hn(t,\bm{k})
\end{bmatrix}
\\
&=
\begin{bmatrix}
(i\bar{\Omega}^{(0)}_{\phi \phi }-L^{(0)}_{\phi\phi})\stiffness k^2& (1-L^{(2)}_{\phi n} k^2)\chi^{nn} \\
(-1-L^{(2)}_{n\phi} k^2)\stiffness k^2& -L^{(2)}_{nn}\chi^{nn} k^2\\
\end{bmatrix}
\begin{bmatrix}
\hphi(t,\bm{k})\\
\delta\hn(t,\bm{k})
\end{bmatrix}.
\end{split}
\end{equation}
Then, the equation of motion for $\hphi(\omega,\bm{k})$ reads in the leading order
\begin{equation}
\begin{split}
[(-i\omega+\Gamma k^2)^2+v^2k^2]\hphi(\omega,\bm{k})=0,
\end{split}
\end{equation}
where $\Gamma=(\chi^{nn}L_{nn}^{(2)}+L_{\phi\phi}^{(0)}\stiffness)/2$, and the higher-order terms of $k$ were dropped.
The formal solution of this is $\omega = \pm \sqrt{v^2}k- i\Gamma k^2$, so that the eigenvalues of the matrix $\pm \sqrt{v^2}k- i\Gamma k^2$ give the dispersion relations,
$\omega = \pm v_ik- i\Gamma_i k^2$. 
The number of type-A NG modes is equal to the number of type-A charges, $N_\typeA=\NBS-\rank\average{ [i\hQ_a,\hQ_b]}$.
At small $k\ll v_i/\Gamma_i$, the real parts are always larger than the imaginary parts, so that the spectra of type-A NG modes become sharper as $k$ decreases.

\subsubsection{Dispersion relations of type-B NG modes}
Here, we derive the dispersion relations of type-B NG modes.
Since $ \deltaB\hn_{\bar{a}}^\typeB(t,\bm{x})$ has the same quantum number under the unbroken symmetry with that of $\hphi^{\typeB{\bar{a}}}(t,\bm{x})$, they mix in the free energy.
Some of them become gapped modes, as will be discussed in the next sub-subsection.
The examples of type-B NG modes are spin waves in the ferromagnetic phase of the Heisenberg model, and the NG modes in the Kaon condensed color super conducting phase~\cite{Miransky:2001tw,Schafer:2001bq}.   In the former example, the only charge densities are the dynamical degrees of freedom, so that they have no gapped partners,
while in the latter example, type-A and type-B NG modes coexist, and a type-B NG mode has a gapped partner.
When the scale of the gap is comparable or larger than the typical energy scale of the system, one need not to employ the both $\deltaB\hn_{\bar{a}}^\typeB(t,\bm{x})$ and $\hphi^{\typeB{\bar{a}}}(t,\bm{x})$ as the slow variables.  Here, we only employ $ \deltaB\hn_{\bar{a}}^\typeB(t,\bm{x})$ as slow variables.

It is useful to decompose $\deltaB \hn_{\bar{a}}^\typeB(t,\bm{x})$ into $ \deltaB{\hn}_{+\bar{a}}^\typeB(t,\bm{x})\equiv \deltaB{\hn}_{\bar{a}}^\typeB(t,\bm{x})$  and $ \deltaB{\hn}_{-\bar{a}}^\typeB(t,\bm{x})\equiv \deltaB{\hn}_{\bar{a}+N_\typeB}^\typeB(t,\bm{x})$.
In this decomposition, the frequency matrix becomes
\begin{equation}
i\Omega^{(0)}_{ n^\typeB_{+\bar{a}}  n^\typeB_{-\bar{b}}}=-i\Omega^{(0)}_{ n^\typeB_{-\bar{b}}  n^\typeB_{+\bar{a}}}= -\frac{1}{\beta}\average{[ i\deltaB\hQ^\typeB_{+\bar{a}}, \deltaB\hn^\typeB_{-\bar{b}}(\bm{x})]}= -\frac{1}{\beta} \lambda_{\bar{a}} \delta_{\bar{a}\bar{b}},
\end{equation}
in the leading order, and others are zero.
In the following of this sub-subsection, we omit the index `B' to avoid complexed  indices and use matrix notation.
Since $ \hn_{\pm}$ are elastic variables, the inverse metric is proportional to $k^2$,
\begin{equation}
g^{n_\pm n_\pm}(\bm{k}) = \beta\stiffness_{\pm} k^2,
\end{equation}
so that $g^{(0)n_{\pm}n_{\pm}}=g^{(0)n_{\pm}n_{\pm}}=0$ and $g^{(2)n_{\pm}n_{\pm}}=\beta \stiffness_{\pm}$ are obtained.
The memory functions for $ \hn_{\pm\bar{a}}^\typeB(t,\bm{x})$ vanish at $\bm{k}=\bm{0}$ because of the conservation law, so that all $i\delta\Omega^{(0)}_{ n^\typeB  n^\typeB}$ and $L^{(0)}_{ n^\typeB  n^\typeB}$ do not appear in the equations of motion.
Therefore, for the type-B NG modes, the equations of motion are given as
\begin{equation}
\begin{split}
\partial_0 
\begin{bmatrix}
\deltaB\hn_+(t,\bm{k})\\
\deltaB\hn_-(t,\bm{k})
\end{bmatrix}
&=
\begin{bmatrix}
-L_{n_+n_+}^{(2)}k^2 &-\Lambda\\
\Lambda &-{L}_{n_-n_-}^{(2)}k^2
\end{bmatrix}
\begin{bmatrix}
\stiffness_{+} k^2&0\\
0 &{\stiffness}_{-} k^2
\end{bmatrix}
\begin{bmatrix}
\deltaB\hn_+(t,\bm{k})\\
\deltaB\hn_-(t,\bm{k})
\end{bmatrix}
\\
&=
\begin{bmatrix}
-L_{n_+n_+}^{(2)}\stiffness_{+} k^4 &-\Lambda {\stiffness}_{-} k^2\\
\Lambda{\stiffness}_{+} k^2 &-{L}_{n_-n_-}^{(2)}{\stiffness}_{-} k^4
\end{bmatrix}
\begin{bmatrix}
\deltaB\hn_+(t,\bm{k})\\
\deltaB\hn_-(t,\bm{k})
\end{bmatrix}.
\end{split}
\end{equation}
In the leading order, the equation of motion for $\deltaB\hn_+(\omega,\bm{k})$ becomes  
\begin{equation}
\begin{split}
[  (-i\omega +\Gamma k^4)^2 + v^2 k^4]\deltaB\hn_+(\omega,\bm{k})=0, \label{eq:EOMtypeB}
\end{split}
\end{equation}
where $v^2=\Lambda\stiffness_{-}\Lambda{\stiffness}_+$, $\Gamma=(L_{n_+n_+}\stiffness_++\Lambda\rho_-{L}_{n_-n_-}\Lambda^{-1 })/2$.
The dispersion relations become $\omega=\pm v_ik^2  -i\Gamma_i k^4$,
which are the eigenvalues of the matrix solution of Eq.~\eqref{eq:EOMtypeB}, i.e., $\omega= \pm \sqrt{v^2}k^2 -i
k^4\Gamma$.
The Poisson bracket is given as
\begin{equation}
\begin{split}
\{\deltaB n_{\bar{a}}(\bm{x}),\deltaB n_{\bar{b}}(\bm{x}')\}_P = -\lambda_{\bar{a}}\delta_{\bar{a}\bar{b}}\delta^{(3)}(\bm{x}-\bm{x}').
\end{split}
\end{equation}
In contrast to the case of type-A NG modes, the effective free-energy has only derivative terms,
\begin{equation}
\begin{split}
F[n] = \int d^3x\Bigl(\frac{1}{2}\stiffness_{+}^{\bar{a}\bar{b}} \partial_i \deltaB n_{+\bar{a}}(\bm{x}) \partial_i \deltaB n_{+\bar{b}}(\bm{x})
+\frac{1}{2}{\stiffness}_-^{\bar{a}\bar{b}} \partial_i \deltaB n_{-\bar{a}}(\bm{x}) \partial_i\deltaB n_{-\bar{b}}(\bm{x})
\Bigr).
\end{split}
\end{equation}
These correspond to Eqs.~\eqref{eq:PoissonBracketTypeB} and \eqref{eq:ElasticFreenergyTypeB} in Sec.~\ref{sec:Examples}, respectively.
The number of type-B NG modes coincides with the number of canonical pairs, $N_\typeB=\rank \Lambda= \rank\average{[i\hQ_a,\hQ_b]}/2$.

\subsubsection{Gapped partners of type-B NG modes} \label{sec:typeB2DispersionRelation}
Next, we discuss the gapped partners of type-B NG modes by explicitly taking into account $\hphi^{\typeB\bar{a}}(t,\bm{x})$.
The number of independent type-B NG fields is equal to $\rank \average{[i\hQ^\typeB_{\bar{a}},\hphi^{\typeB \bar{b}}(\bm{x})]}=\rank \average{[i\hQ_a,\hphi_i(\bm{x})]}-N_\typeA$. 
The type-B NG fields will correspond to the gapped modes. If the scale of the order parameter $W_{\bar{a}\bar{b}}\sim n^0$ is the same order as the typical energy scale of non-NG modes, $\varLambda_{\text{UV}}$, they are not dynamical variables in the low-energy region. One may integrate $\hphi^{\typeB\bar{a}}(t,\bm{x})$ out explicitly, and obtain the same result as that in the previous sub-subsection. 
On the other hand,  if the scale of $W_{\bar{a}\bar{b}}$ is much smaller than $\varLambda_{\text{UV}}$, the gapped partners play roles of low-energy degrees of freedom, which are called ``almost NG modes''~\cite{Kapustin:2012cr,Hidaka:2012ym,Nicolis:2013sga,Gongyo:2014sra}. 
Here, we show that the equations of motion for $\hphi^{\typeB\bar{a}}(t,\bm{x})$ and $\deltaB\hn_{\bar{a}}^{\typeB}(t,\bm{x})$ contain the both gapped and gapless modes.
In the following, we again omit the index `B' and use the matrix notation.

To see the gapped degrees of freedom, we take the low-momentum limit $\bm{k}\to\bm{0}$. 
The frequency matrices,
$i\Omega^{(0)}_{n\phi} $ and $i\Omega^{(0)}_{n\phi} $, are
\begin{align}
i\Omega^{(0)}_{n\phi} =-\frac{1}{\beta}\average{[i\deltaB\hQ,\hphi(\bm{x})]}&=-\frac{1}{\beta}\mathcal{Q},\\
i\Omega^{(0)}_{nn} =-\frac{1}{\beta}\average{[i\deltaB\hQ,\delta\hn(\bm{x})]}&=-\frac{1}{\beta}W.
\end{align}
At $\bm{k}=\bm{0}$, the memory functions $K_{nn}$, $K_{\phi n}$ and $K_{n\phi }$ vanish, and the only $K_{\phi\phi}$ contributes to the equations of motion.
We assume the renormalized frequency matrix, $i\Omega^{(0)}_{lm}$ is a regular matrix. If this is not the case, some of degrees of freedom are not independent in the sense of canonical variables. 
For $\phi$ sector, we define $M \equiv \beta(i\bar{\Omega}^{(0)}_{\phi\phi} -L^{(0)}_{\phi\phi}- \delta Z^{(0)}_{\phi\phi} \partial_0)$.
Then, the equations of motion read
\begin{equation}
\begin{split}
\partial_0
\begin{bmatrix}
\hphi(t,\bm{k}=\bm{0})\\
\deltaB\hn(t,\bm{k}=\bm{0})
\end{bmatrix}
&=
\begin{bmatrix}
M& \mathcal{Q}\\
-\mathcal{Q}&-W\\
\end{bmatrix}
\begin{bmatrix}
\chi^{\phi\phi} & \chi^{\phi n}\\
\chi^{n\phi}&\chi^{n n}
\end{bmatrix}
\begin{bmatrix}
\hphi(t,\bm{k}=\bm{0})\\
\deltaB\hn(t,\bm{k}=\bm{0})
\end{bmatrix}
\\
&=
\begin{bmatrix}
RW\mathcal{Q} &
R
\\
0& 0
\end{bmatrix}
\begin{bmatrix}
\hphi(t,\bm{k}=\bm{0})\\
\deltaB\hn(t,\bm{k}=\bm{0})
\end{bmatrix},
\end{split}
\end{equation}
where we used $-\mathcal{Q}\chi^{\phi\phi}-W\chi^{n\phi}=-\mathcal{Q}\chi^{\phi n}-W\chi^{nn}=0$,
which can be obtained from Eq.~\eqref{eq:FlattnessOfFreenergy}.
These express the charge conservation $\partial_0\deltaB\hn(t,\bm{k}=\bm{0})=\partial_0\deltaB\hQ=0$.
We defined $R\equiv M\chi^{\phi n}+\mathcal{Q}\chi^{n n}= \mathcal{Q}(1-MW)\chi^{nn}$ 
and $M\chi^{\phi\phi}+\mathcal{Q}\chi^{n\phi }= RW\mathcal{Q}$ (Note that $\mathcal{Q}M=M$ holds.)
The equation of motion for $\hphi$ is given as
\begin{equation}
\begin{split}
(\partial_0 -RW\mathcal{Q})\partial_0 \hphi(t,\bm{k}=\bm{0}) =0.
\end{split}
\end{equation}
We are interested in the gapped modes that satisfy $\partial_0\hphi(t,\bm{k}=\bm{0})\neq0$.
When $W$ is much smaller than $\varLambda_{\text{UV}}$,  $RW\mathcal{Q}\simeq \mathcal{Q}\chi^{nn}W\mathcal{Q}$.
The eigenvalues of $i\mathcal{Q}\chi^{nn}W\mathcal{Q}$ give the gaps of modes, which are proportional to  $W\sim n^0$.
The number of gapped modes coincides with
\begin{equation}
\begin{split}
N_\text{gapped}=\frac{1}{2}\rank\mathcal{Q}=\frac{1}{2}\bigl( \rank \average{[i\hQ_a,\hphi_i(\bm{x})]}-N_\typeA\bigr).
\end{split}
\end{equation}

Let us briefly check that the other modes are type-II NG modes.
For this purpose, we consider a perturbation by small momentum $k^2$.
\begin{equation}
\begin{split}
\partial_0
\begin{bmatrix}
\hphi(t,\bm{k})\\
\deltaB\hn(t,\bm{k})
\end{bmatrix}
&=
\begin{bmatrix}
RW\mathcal{Q} &
R
\\
g_1k^2& g_2k^2
\end{bmatrix}
\begin{bmatrix}
\hphi(t,\bm{k})\\
\deltaB\hn(t,\bm{k}) 
\end{bmatrix},
\end{split}
\end{equation}
where $g_1$ and $g_2$ are coefficient matrices, and  we neglected the corrections to $RW\mathcal{Q}$ and $R$ that are higher orders.
From the equation of motion for $\partial_0\hphi(t,\bm{k})$, we obtain $\hphi(t,\bm{k})\simeq -\mathcal{Q}W^{-1}\mathcal{Q}\delta\hn(t,\bm{k})$, and then, the equation for $\delta\hn(t,\bm{x})$ becomes
\begin{equation}
\begin{split}
\partial_0 \delta\hn(t,\bm{k}) = g_1k^2 \hphi(t,\bm{k}) +g_2k^2\delta\hn(t,\bm{k}) \simeq (-g_1 \mathcal{Q}W^{-1}\mathcal{Q} + g_2)k^2 \deltaB\hn(t,\bm{k}).
\end{split}
\end{equation}
This is nothing but the equation of motion for type-II NG modes.

\subsubsection{Explicit breaking and gap formula}
In this sub-subsection, we discuss the effect of a small explicit breaking term on the NG modes.
For this purpose, we add $\epsilon\hV$ with a small expansion parameter $\epsilon$ into the Hamiltonian, where 
the symmetry is approximate and the NG modes are no longer gapless but gapped, which  are called pseudo-NG modes~\cite{WeinbergText}.
We assume that the explicit breaking term does not change the symmetry breaking pattern.
In other words, the vacuum alignment condition is satisfied at $\epsilon\to0$. 

The explicit breaking term modifies  Eq.~\eqref{eq:STIdentity} to 
\begin{equation}
\begin{split}
\int d^3x'\frac{\delta F}{\delta \Phi_l(\bm{x}')}\OP_{al}(\bm{x}';J)=\average{[i\hQ_a,\epsilon\hV]}_J. \label{eq:STIdentityBreaking}
\end{split}
\end{equation}
The functional derivative of Eq.~\eqref{eq:STIdentityBreaking} with respect to $\Phi_m(\bm{x})$ leads to 
\begin{equation}
\begin{split}
\int d^3x' \chi^{ml}(\bm{x}-\bm{x}')\OP_{al}(\bm{x}') =\frac{\delta}{\delta \Phi_m(\bm{x})}\average{[i\hQ_a,\epsilon\hV]}_{J=0}\equiv \epsilon w_{a}^{~m}.
\label{eq:explictBreaking}
\end{split}
\end{equation}
We note that the $J(\bm{x})=0$ limit is taken after the functional derivative.
We are interested in the gap of the pseudo-NG modes, so that let us take the $\bm{k}=\bm{0}$ limit.
The equation of motion for $\delta\hn_a(t,\bm{k}=\bm{0})$ becomes
\begin{equation}
\begin{split}
\partial_0 \delta\hn_a(t,\bm{k}=\bm{0}) &= -\epsilon w_{a}^{~m}\hPhi_m(t,\bm{k}=\bm{0}) + \mathcal{O}(\epsilon^2)\\
&=  -\epsilon w_{a}^{~\phi_i}\hphi_i(t,\bm{k}=\bm{0}) -\epsilon  w_{a}^{~n_b} \delta\hn_b(t,\bm{k}=\bm{0}) + \mathcal{O}(\epsilon^2).
\label{eq:PCAC}
\end{split}
\end{equation}
For type-A NG modes, the equations of motion can be written as, in the leading order of $\epsilon$,
\begin{align}
\partial_0\hphi^{\typeA a }(t,\bm{k}=\bm{0}) &= \chi^{n_an_b} \delta\hn^{\typeA}_b(t,\bm{k}=\bm{0}) + \mathcal{O}(\epsilon),\\
\partial_0 \delta\hn^{\typeA}_a(t,\bm{k}=\bm{0}) &= -\epsilon w_{a}^{~\phi^b}\hphi^{\typeA b}(t,\bm{k}=\bm{0})-\epsilon  w_{a}^{~ n_b}\delta\hn^{\typeA}_b(t,\bm{k}=\bm{0}) +\mathcal{O}(\epsilon^2).
\end{align}
Then, the equation of motion for $\hphi^{\typeA a }(t,\bm{k}=\bm{0})$ reads
\begin{equation}
\begin{split}
\partial_0^2\hphi^{\typeA a}(t,\bm{k}=\bm{0}) = -[m_\typeA^{ 2}]^a_{~b}\hphi^{\typeA b}(t,\bm{k}=\bm{0}) +\mathcal{O}(\epsilon^2),
\end{split}
\end{equation}
where $m_\typeA^2$ is the gap matrix,
\begin{equation}
\begin{split}
[m_\typeA^{2}]^a_{~b} \equiv  \epsilon \chi^{n_an_c} w_c^{~\phi^b}. \label{eq:massmatrix}
\end{split}
\end{equation}
This  relation corresponds to the generalized Gell-Mann--Oakes--Renner relation~\cite{Son:2001ff,*Son:2002ci}. 
In QCD, these parameters correspond to
the quark mass $\epsilon=m_q$, the chiral condensate $w=-\average{\hat{\bar{\psi}}\hat{\psi}}$, and the pion decay constant $F_\pi^{-2}=\chi^{nn}$, respectively.
Thus, the pion mass is given by $m_\pi^2=-m_q\average{\hat{\bar{\psi}}\hat{\psi}}/F_\pi^2$~\cite{GellMann:1968rz}.
We note that we implicitly assumed $w_{a}^{~B}=\mathcal{O}(1)$ to derive the gap formula, in which the gap is of order $\sqrt{\epsilon}$.
 If $w_{a}^{~B}=\mathcal{O}(\epsilon)$, other terms can contribute to Eq.~\eqref{eq:massmatrix}.
In fact, this is the case for the chiral symmetry breaking in the color flavor locking phase, in which the mass of NG modes are of order $\sqrt{m_q^2}$~\cite{Son:1999cm,*Son:2000tu}.

On the other hand, for type-B NG modes,
the equation of motion at $\bm{k}=\bm{0}$ reads
\begin{equation}
\begin{split}
\partial_0  \deltaB\hn^\typeB_{\bar{a}}(t,\bm{k}=\bm{0}) = -\epsilon  w_{\bar{a}}^{ n_{\bar{b}}}\deltaB\hn^{\typeB}_{\bar{b}}(t,\bm{k}=\bm{0}). \label{eq:GapEquationTypeB}
\end{split}
\end{equation}
Thus, the gap matrix is obtained as
\begin{equation}
\begin{split}
[m_\typeB^2]_{\bar{a}}^{~\bar{b}}= -\epsilon^2 w_{\bar{a}}^{~n_{\bar{c}}} w_{\bar{c}}^{~n_{\bar{b}}}.
\end{split}
\end{equation}
In contrast to the case of type-A pseudo-NG modes, the gap is of order $\epsilon$.

If the explicit breaking term is proportional to a linear combination of charge operators, $\hV=-\mu^B\hQ_B$, we have $[i\hQ_{\bar{a}},\hV]=-[i\hQ_{\bar{a}},\mu^B\hQ_B]=\mu^B f_{\bar{a}B}^{~~C}\hQ_{C}$.
Therefore, $w_{\bar{a}}^{~ n_{\bar{c}}}= \mu^Bf_{\bar{a}B}^{~~\bar{c}}$, and the mass matrix becomes
\begin{equation}
\begin{split}
[m^2]_{\bar{a}}^{~\bar{b}}= -f_{\bar{a}A}^{~~\bar{c}}f_{\bar{c}B}^{~~\bar{b}}\mu^A\mu^B.
\end{split}
\end{equation}
Note that this gap is exact and the modes do not dissipate at $\bm{k}=\bm{0}$.
This is because the Heisenberg equation for $\hQ_a(t)$  satisfies
$\partial_0\hQ_A(t) =i[\hH-\mu^B\hQ_B, \hQ_A(t)] = \mu^B f_{BA}^{~~C}\hQ_C(t)$,
which corresponds to Eq.~\eqref{eq:GapEquationTypeB}.

In this paper, we considered the cases in which the chemical potential term commutes with other charges, i.e., $[\mu \hN, \hQ_A]=0$.
If one uses $\mu^B\hQ_B$ as the chemical potential term, in which $[\hQ_A,\mu^B\hQ_B]=\mu^B if_{AB}^{~~C}\hQ_{C}$, one obtains Eq.~\eqref{eq:explictBreaking} with $w_{a}^{~n_c}=\mu^Bf_{aB}^{~~c}$.
However, $w_{a}^{~n_c}$ exactly cancels with the chemical potential term in the streaming term of Eq.~\eqref{eq:StreamingChemical},
and we obtain $\partial_0 \delta\hn_a^\typeA(t,\bm{k}=\bm{0})=0$, which is nothing but the charge conservation law.
Therefore, the dispersion relation of the NG mode obtained from the pole of the response function has no gap.
However, there is an ambiguity of the definition of the gap.
Since we use the Hamiltonian as the time evolution operator, the obtained dispersion relation corresponds to  the 
difference of the energy from the thermal state without the chemical potential energy. 
If one wants to measure the difference of the free energy from the thermal state, one may 
add the chemical potential energy $\mu^Bf_{aB}^{~~c}$ into the dispersion relation.
In our formulation, such a dispersion relation can be obtained by replacing the time evolution operator $\hH$ with $\hH-\mu^B\hQ_B$,
which is equivalent to adding the explicit breaking term $\hV=-\mu^B\hQ_B$ into the Hamiltonian.
At zero temperature and finite density, such gapped modes are discussed in Refs.~\cite{Nicolis:2012vf,Nicolis:2013sga,Watanabe:2013uya}.

\subsubsection{Mixing with hydrodynamic modes}\label{sec:HydrodynamicModes}
So far, we have not taken into account the effect of hydrodynamic modes (diffusion and acoustic sound modes) to the dispersion relations of NG modes.
In general, we need to consider the Langevin equations of all charge densities including energy and momentum densities.
If a NG mode belongs to nonsinglet representation under the unbroken symmetry, it cannot couple to the hydrodynamic modes
because the energy and momentum densities are singlet. 
Conversely, if the NG mode belongs to the singlet representation under the unbroken symmetry, it does.
It is known that in the superfluid phase, the type-A NG mode associated with spontaneous breaking of particle number symmetry 
couples to the sound mode of the normal fluid component, and this mixing causes the second sound mode.
The mixing does not change the powers of momentum in the dispersion relation, but the velocity is modified~\cite{Chaikin}.
In order to derive this modification from our formulation, we neglect the dissipation terms. 
For simplicity, we consider the one type-A mode, whose local operator and charge density are denoted by $\hphi(\bm{x})$ and $\hn(\bm{x})$, respectively.
We consider a relativistic fluid with energy and momentum densities, $\hat{e}(\bm{x})$ and $\hat{p}^i(\bm{x})$.
The equation of motion for the energy density is
\begin{equation}
\begin{split}
\partial_0 \delta\hat{e}(t,\bm{x})  =-\partial_i \hat{p}^{i}(t,\bm{x}), \label{eq:EnergyConservation}
\end{split}
\end{equation}
which is nothing but the energy conservation law.
Since we chose $\hat{p}^i(\bm{x})$ as slow variables, this equation is exact; no other variables couple to the energy density~\cite{Minami:2012hs}.

In order to determine  the equation of motion for $\hat{p}^i(t,\bm{x})$,  we evaluate  the frequency matrices:
\begin{align}
i\Omega_{p^i e}(\bm{k})&= -\frac{1}{\beta}\int d^3x e^{-i\bm{k}\cdot\bm{x}}\average{[i\hat{p}^{i}(\bm{x}), \delta \hat{e}(\bm{0}) ]}\notag\\ 
&=-\frac{1}{\beta}ik^j \average{\delta_{ij}\hat{e}(\bm{0})+\hT_{ij}(\bm{0})}+\mathcal{O}(k^3)
=-\frac{1}{\beta}ik^ih+\mathcal{O}(k^3),  \label{eq:pie}\\
i\Omega_{p^i n}(\bm{k})&=-\frac{1}{\beta}\int d^3x e^{-i\bm{k}\cdot\bm{x}}\average{[i\hat{p}^{i}(\bm{x}), \delta \hn(\bm{0}) ]}\notag\\
& =-\frac{1}{\beta} ik^i \average{\hn(\bm{0})}+\mathcal{O}(k^3)= -\frac{1}{\beta}ik^i n +\mathcal{O}(k^3) \label{eq:pin},\\
i\Omega_{p^i \phi}(\bm{k})&=-\frac{1}{\beta}\int d^3x e^{-i\bm{k}\cdot\bm{x}}\average{[i\hat{p}^{i}(\bm{x}), \hphi(\bm{0}) ]}\notag\\
& =0,
\label{eq:pphi}
\end{align}
where $\hT_{ij}(\bm{x})$ is the momentum tensor, 
the charge density $n\equiv\average{\hn(\bm{x})}$, and the enthalpy $h=e+p$ with the energy density $e\equiv \average{\hat{e}(\bm{x})}$ and the pressure $p\delta_{ij} \equiv \average{\hT_{ij}(\bm{x})}$. $\Omega_{p^i \phi}(\bm{k})$ vanishes due to time-reversal symmetry.
From Eqs.~\eqref{eq:pie} to \eqref{eq:pphi}, the leading order terms vanish:
$ i\Omega^{(0)e}_{p^i}= i\Omega^{(0)n}_{p^i}= i\Omega^{(0)p^i}_{n}=0$, and the next leading order terms are
\begin{equation}
\begin{split}
 i\Omega^{(1)e}_{p^i} &= h \chi^{ee} +n\chi^{ne}\equiv P_e,\\
 i\Omega^{(1)n}_{p^i}&=h\chi^{en}+n\chi^{nn}\equiv P_n,\\
 i\Omega^{(1)p^i}_{n}&=n\chi^{p^ip^i} =\frac{n}{h},
\end{split}
\end{equation}
where $\chi^{ee}$, $\chi^{ne}$, $\chi^{nn}$, and $\chi^{p^ip^i}$ are inverse susceptibilities, respectively.
Here, we used $h=\chi_{p^ip^i}$~\cite{Minami:2012hs}.
In the streaming term of $\hphi(t,\bm{x})$, in addition to $ i\Omega^{(0)n}_{\phi}=\chi^{nn}$, the mixing term with energy density, $i\Omega^{(0)e}_{\phi}=\chi^{ne}$, appears.
Then,  the equations of motion can be written as, in the leading order,
\begin{align}
\partial_0 \hphi(t,\bm{x})&= \chi^{n n}\delta \hn(t,\bm{x}) +  \chi^{n e}\delta\hat{e}(t,\bm{x}), \label{eq:EOMphi}\\
\partial_0 \hat{p}^i(t,\bm{x})& =  P_n\partial^i \delta\hat{n}(t,\bm{x}) + P_e\partial^i \delta \hat{e}(t,\bm{x}),\\
\partial_0 \delta\hn(t,\bm{x}) &= \rho\partial_i^2 \hphi(t,\bm{x}) - \frac{n}{h}\partial_i\hat{p}^i(t,\bm{x}). \label{eq:EOMn2}
\end{align}
If $\hphi(t,\bm{x})$ is absent, these equations reproduce the linearized hydrodynamic equations for a normal fluid~\cite{Minami:2012hs}.
From Eqs.~\eqref{eq:EnergyConservation} and \eqref{eq:EOMphi} to \eqref{eq:EOMn2}, the equations of motion for $\delta \hn(t,\bm{x})$ and $\delta \hat{e}(t,\bm{x})$ read
\begin{equation}
\partial_0^2
\begin{bmatrix}
\delta \hn(t,\bm{x}) \\
\delta\hat{e}(t,\bm{x}) \\
\end{bmatrix}
=
\begin{bmatrix}
c_{11} & c_{12} \\
c_{21} & c_{22}
\end{bmatrix}
\partial_i^2
\begin{bmatrix}
\delta \hn(t,\bm{x}) \\
\delta\hat{e}(t,\bm{x}) \\
\end{bmatrix},
\end{equation}
where $c_{11}\equiv nP_n/h+\rho\chi^{nn}$, $c_{12}\equiv nP_e/h+\rho\chi^{n e}$, $c_{21}\equiv P_n$, $c_{22}\equiv P_e$, respectively.
The dispersion relations are given as $\omega =\pm k\sqrt{(c_{11}+c_{22}\pm\sqrt{(c_{11}-c_{22})^2+4c_{12} c_{21}})/2} $, which correspond to the first and second sounds, respectively.
If one takes into account the dissipation effects, one may add the shear and bulk viscosity terms into the equations for $\delta\hat{p}^i(t,\bm{x})$,
in addition to dissipation  terms for $\hphi(t,\bm{x})$ and $\delta\hn(t,\bm{x})$, which give the $k^2$ contributions to the imaginary parts of the dispersion relations.

Similarly, hydrodynamic modes modify the dispersion relations of type-B NG modes when $\average{\hn_{\bar{a}}^\typeB(\bm{x})}\neq0$.
However, they do not change the powers of the dispersion relation as in the case of type-A NG modes
 because the mixing term between the hydrodynamic and NG modes necessarily contains the spatial derivatives.

\section{Summary}\label{sec:summary}
We have discussed the dispersion relations of NG modes associated with spontaneous breaking of internal symmetries at finite temperature and/or density. 
The dispersion relations for type-A and type-B are given by 
$\omega_\typeA= v_\typeA k-i\Gamma_\typeA k^2$ and $\omega_B= v_\typeB k^2-i\Gamma_\typeB k^4$, respectively.
The coefficients, $v_{\typeA,\typeB}$ and $\Gamma_{\typeA,\typeB}$,  depend on the details of theory.  
At small $k$, the imaginary parts are smaller than the real parts, so that  the both of type-A and type-B NG modes can propagate a long distance.
This does not hold for spontaneous breaking of spacetime symmetries.
For example,  the dispersion relation of NG mode in a nematic crystal phase becomes  $\omega = vk^2-i\Gamma k^2$~\cite{Hosino:1982},
 where the real and imaginary parts are the same order; in particular, the parameter $v$ depends on the temperature, and it vanishes at some temperature, i.e., the mode is overdamping.

Here, we summarize the relation between the number of broken symmetries, order parameters,  type-A, type-B, and gapped modes:
\begin{align}
N_\typeB &= \frac{1}{2}\rank \average{[i\hQ_a,\hQ_b]},  \label{eq:NtypeB}\\
 N_\typeA &=\NBS -2N_\typeB, \label{eq:NtypeA}\\
N_\text{gapped} &= \frac{1}{2}\Bigl(\rank \average{[i\hQ_a,\hphi_i(\bm{x})]}-N_\typeA\Bigr).
\end{align}
Equations~\eqref{eq:NtypeB} and \eqref{eq:NtypeA} lead to Eq.~\eqref{eq:NGRelation}.
The number of gapped modes can also read
\begin{equation}
N_\text{gapped} = \frac{1}{2}\Bigl(\rank \average{[i\hQ_a,\hphi_i(\bm{x})]} +\rank\average{[i\hQ_a,\hQ_b]}-\NBS\Bigr).
\end{equation}
A similar relation was obtained using the effective Lagrangian approach, where the rank of $\average{[i\hQ_a,\hphi_i(\bm{x})]}$ corresponds to that of coefficient matrix $\bar{g}_{ab}$ in the second time-derivative term~\cite{Gongyo:2014sra}.
 
 We also derived the gap formula when the symmetry is explicitly broken.  This is the generalization of Gell-Mann--Oakes--Renner relation in QCD to 
  finite temperature and/or density.
 For type-A NG modes, the gap matrix is proportional to the square root of the explicit breaking parameter $\epsilon$,  while for type-B NG modes, it is linearly proportional to $\epsilon$.

In this paper, we focused on the dispersion relations of NG modes associated with spontaneous breaking of  internal symmetries. 
It will be interesting to generalize our work to the case of spontaneous breaking of spacetime symmetries.

\acknowledgements
We thank  A.~Beekman, S.~Gongyo, Y.~Hirono, S.~Karasawa, and Y.~Tanizaki for useful discussions.  
We also thank the Yukawa Institute for Theoretical Physics at Kyoto University.
Discussion during the YITP workshop YITP-W-14-02 on ``Higgs Modes in Condensed Matter and Quantum gases"
were useful to complete this work. 
T.~H.~was supported by JSPS Research Fellowships for Young Scientists.
Y.~H. was partially supported by JSPS KAKENHI Grants Numbers 24740184. 
This work was also partially supported  by the RIKEN iTHES Project.

\appendix
\section{Relation between the kubo response and retarded functions}\label{sec:RelationBetweenResponseFunctions}
In the linear response theory, there are two types of response functions: One is the retarded Green function that appears in the perturbation by external fields such as an electric field. The other is the Kubo response function that appears in a relaxation process from a nonequilibrium state.
These two functions are not independent, and the poles of the response functions coincide with each other.
The retarded Green function is defined by
\begin{equation}
\begin{split}
G_{n}^{Rm}(t)\equiv i\theta(t) \average{[\hA_n(t), \hA^{m\dag}]}.
\end{split}
\label{eq:retarded}
\end{equation}
In order to see the relation between them, we perform the Laplace transformation of Eq.~\eqref{eq:retarded},
\begin{equation}
\begin{split}
G_{n}^{Rm}(z)= i\int_0^\infty dt e^{-zt} \average{[\hA_n(t), \hA^{m\dag}]}.
\end{split}
\end{equation}
Using the same techniques used in Eq.~\eqref{eq:streaming2}, we have
\begin{equation}
\begin{split}
i\average{[\hA_n(t), \hA^{m\dag}]} =-\beta(\partial_0 \delta_n^{~l}-i\mu q_n^{~l}) \innerProd{  \hA_l(t), \hA^{m}}.
\end{split}
\end{equation}
Then, the retarded Green's function becomes
\begin{equation}
\begin{split}
 G_{n}^{Rm}(z)&=-\beta \int_0^\infty dt e^{-tz}(\partial_0 \delta_n^{~l}-i\mu q_n^{~l})  \innerProd{ \hA_l(t),\hA^{m}}\\
 &= -\beta \bigl((z \delta_n^{~l}-i\mu q_n^{~l}) {G_l}^m(z)-{\delta_n}^m\bigr)\\
& =-\beta{\left[\frac{-i\mu q+i\varOmega-\memory(z)}{z-i\varOmega+\memory(z) }\right]_n}^m.
\end{split}
\end{equation}
In the last line, we used matrix notation. We also used ${G_n}^m(t=0)={\delta_n}^m$. 
Therefore, the poles of $G_{n}^{Rm}(z)$ coincide with those of  ${G_{n}}^m(z)$ in the complex $z$ plane.

\bibliography{paper}
\end{document}